\documentclass[useAMS,usenatbib]{mn2e}

\usepackage{graphicx}
\usepackage{graphics}
\usepackage{color}
\usepackage{url}
\usepackage[caption=false]{subfig}
\usepackage{verbatim}
\usepackage{array}
\usepackage{amssymb}
\usepackage{amsmath}
\usepackage{longtable}
\usepackage{pbox}

\newcommand{\name}{ASASSN-14li}
\newcommand{\galname}{PGC 043234}
\newcommand{\swift}{{\it Swift}}
\newcommand{\msun}{\ensuremath{\rm{M}_\odot}}
\newcommand{\lsun}{\ensuremath{\rm{L}_\odot}}

\newcommand{\ion}[2]{{#1~\uppercase \expandafter{\romannumeral #2}}}
\newcommand{\edit}[1]{\textcolor{black}{#1}}
\newcolumntype{C}[1]{>{\centering\arraybackslash}m{#1}}
\newcolumntype{L}[1]{>{\arraybackslash}m{#1}}

\begin{document}

\title[The Nearby TDE Candidate ASASSN-14li]{Six Months of Multi-Wavelength Follow-up of the Tidal Disruption Candidate ASASSN-14li and Implied TDE Rates from ASAS-SN}

\author[T.~W.-S.~Holoien et al.]{T.~W.-S.~Holoien$^{1,2,3}$\thanks{e--mail: tholoien@astronomy.ohio-state.edu}, C.~S.~Kochanek$^{1,2}$,  J.~L.~Prieto$^{4,5}$, K.~Z.~Stanek$^{1,2}$,
\newauthor
Subo~Dong$^{6}$, B.~J.~Shappee$^{7,8,9}$, D.~Grupe$^{10}$, J.~S.~Brown$^{1}$, U.~Basu$^{1,11}$,
\newauthor
J.~F.~Beacom$^{1,2,12}$, D.~Bersier$^{13}$, J.~Brimacombe$^{14}$, A.~B.~Danilet$^{12}$, E.~Falco$^{15}$,
\newauthor
Z.~Guo$^{6}$, J.~Jose$^{6}$, G.~J.~Herczeg$^{6}$, F.~Long$^{6}$, G.~Pojmanski$^{16}$, G.~V.~Simonian$^{1}$,
\newauthor
D.~M.~Szczygie{\l}$^{16}$, T.~A.~Thompson$^{1,2}$, J.~R.~Thorstensen$^{17}$, R.~M.~Wagner$^{1,18}$, 
\newauthor 
and P.~R.~Wo\'zniak$^{19}$ \\ \\
  $^{1}$ Department of Astronomy, The Ohio State University, 140 West 18th Avenue, Columbus, OH 43210, USA \\
  $^{2}$ Center for Cosmology and AstroParticle Physics (CCAPP), The Ohio State University, 191 W. Woodruff Ave., Columbus, OH 43210, USA \\
  $^{3}$ US Department of Energy Computational Science Graduate Fellow \\
  $^{4}$ N\'ucleo de Astronom\'ia de la Facultad de Ingenier\'ia, Universidad Diego Portales, Av. Ej\'ercito 441, Santiago, Chile \\
  $^{5}$ Millennium Institute of Astrophysics, Santiago, Chile \\
  $^{6}$ Kavli Institute for Astronomy and Astrophysics, Peking University, Yi He Yuan Road 5, Hai Dan District, Beijing, China \\
  $^{7}$ Carnegie Observatories, 813 Santa Barbara Street, Pasadena, CA 91101, USA \\
  $^{8}$ Hubble Fellow\\
  $^{9}$ Carnegie-Princeton Fellow \\
  $^{10}$ Department of Earth and Space Science, Morehead State University, 235 Martindale Dr., Morehead, KY 40351, USA \\
  $^{11}$ Grove City High School, 4665 Hoover Road, Grove City, OH 43123, USA \\
  $^{12}$ Department of Physics, The Ohio State University, 191 W. Woodruff Ave., Columbus, OH 43210, USA \\
  $^{13}$ Astrophysics Research Institute, Liverpool John Moores University, 146 Brownlow Hill, Liverpool L3 5RF, UK \\
  $^{14}$ Coral Towers Observatory, Cairns, Queensland 4870, Australia \\
  $^{15}$ Whipple Observatory, Smithsonian Institution, 670 Mt. Hopkins Road, P. O. Box 6369, Amado, AZ 85645m USA \\
  $^{16}$ Warsaw University Astronomical Observatory, Al. Ujazdowskie 4, 00-478 Warsaw, Poland \\
  $^{17}$ Department of Physics and Astronomy, Dartmouth College, Hanover, NH 03755, USA \\
  $^{18}$ LBT Observatory, University of Arizona, Tucson, AZ 85721-0065 \\
  $^{19}$ Los Alamos National Laboratory, Mail Stop B244, Los Alamos, NM 87545, USA
  }

\maketitle

\begin{abstract}
We present ground-based and {\swift} photometric and spectroscopic observations of the candidate tidal disruption event (TDE) {\name}, found at the center of {\galname} ($d\simeq90$~Mpc) by the All-Sky Automated Survey for SuperNovae (ASAS-SN). The source had a peak bolometric luminosity of $L\simeq10^{44}$~ergs~s$^{-1}$ and a total integrated energy of $E\simeq7\times10^{50}$~ergs radiated over the $\sim6$ months of observations presented. The UV/optical emission of the source is well-fit by a blackbody with roughly constant temperature of $T\sim35,000$~K, while the luminosity declines by roughly a factor of 16 over this time. The optical/UV luminosity decline is broadly consistent with an exponential decline, $L\propto e^{-t/t_0}$, with $t_0\simeq60$~days. {\name} also exhibits soft X-ray emission comparable in luminosity to the optical and UV emission but declining at a slower rate, and the X-ray emission now dominates. Spectra of the source show broad Balmer and helium lines in emission as well as strong blue continuum emission at all epochs. We use the discoveries of {\name} and ASASSN-14ae to estimate the TDE rate implied by ASAS-SN, finding an average rate of \edit{$r \simeq 4.1 \times 10^{-5}~{\rm yr}^{-1}$} per galaxy with a 90\% confidence interval of $(2.2 - 17.0) \times 10^{-5}~{\rm yr}^{-1}$ per galaxy. ASAS-SN found roughly 1 TDE for every 70 Type Ia supernovae in 2014, a rate that is much higher than that of other surveys.
\end{abstract}



\section{Introduction}
\label{sec:intro}

A star orbiting a supermassive black hole (SMBH) can be torn apart if its orbit brings it within the tidal disruption radius of the SMBH where tidal shear forces overpower the self-gravity of the star. In these so-called ``tidal disruption events'' (TDEs), roughly half of the mass of the star may be ejected while the rest of the stellar material is accreted onto the black hole, resulting in a short-lived ($t\la1$~yr) accretion flare \citep[e.g.,][]{lacy82,rees88,phinney89,evans89}. In cases where the central black hole has a mass $M_{BH}\la10^7 {\msun}$, the initial fallback rate is super-Eddington, and the eventual rate at which material returns to pericenter roughly follows a $t^{-5/3}$ power law \citep{evans89,phinney89}. It is commonly assumed that the resulting luminosity of the TDE flare will be proportional to the rate of return of the stellar material to pericenter, but the exact return rates depend on the complex physics associated with the evolution of the accretion stream \citep[e.g.,][]{kochanek94,lodato11,guillochon15,shiokawa15}.

While the accretion of the stellar material powers the TDE flare, direct emission from the disk is only expected to be seen in late phases, with the emission for the majority of the duration of the flare likely dominated by a photosphere formed in the stellar debris \citep{evans89,loeb97,ulmer99,strubbe09}. However, simulations have shown that the source of emission seen by the observer likely depends on the viewing angle of the event \citep{guillochon14}. While sharing some characteristics similar to both supernovae (SNe) and active galactic nuclei (AGN), TDEs are expected to show unique spectral characteristics and light curve evolution that would distinguish them from such transients. Notably, \citet{holoien14b} showed that TDEs exhibit significant long-lasting blue$-$ultraviolet (UV) flux increases compared to supernovae and \citet{arcavi14} found that TDE candidates span a range of H- to He-dominated spectral features. As the light emitted during the TDE may be sensitive to the black hole spin and mass \citep[e.g.,][]{magorrian99,ulmer99,graham01,metzger15}, the detection and study of TDEs may provide a unique method for studying the properties of SMBHs.

A number of candidate TDEs were discovered by UV and X-ray surveys (e.g., NGC5905; \citet{komossa99}, RX J1624+75; \citet{grupe99}, A1795; \citet{donato14}, Swift J164449.3+573451; \citet{burrows11,bloom11,levan11,zauderer11}, Swift J0258.4+0516; \citet{cenko12b}, and GALEX candidates D1-9, D3-13, and D23H-1; \citet{gezari08,gezari09}). Many of these cases did not show correspondingly strong optical emission, and in some cases they were discovered in sparse archival data, limiting their usefulness as probes of TDE physics. In recent years, however, a number of candidates have been found by high-cadence optical surveys, including the All-Sky Automated Survey for Supernovae (ASASSN-14ae, \citealt{holoien14b}), the Palomar Transient Factory (PTF10iya, \citealt{cenko12a}; PTF09ge, PTF09axc, and PTF09djl, \citealt{arcavi14}), Pan-STARRS (PS1-10jh, \citealt{gezari12b}; PS1-11af, \citealt{chornock14}), and the Sloan Digital Sky Survey (TDE 1 and TDE 2, \citealt{velzen11}). These transients typically showed strong UV and optical emission but no associated X-ray emission, and were often better-studied than previous candidates, as they were discovered and followed up by survey projects searching their data in real-time. Optically discovered candidates have provided new opportunities to study these rare transients in greater detail than previously possible, and recent estimates based on optical TDE discoveries put the TDE rate at $\dot{N}_{TDE}=$~(1.5$-$2.0)$^{+2.7}_{-1.3}\times10^{-5}$~yr$^{-1}$ per galaxy \citep{velzen14}. However, there is still tension between this estimate and the rates predicted by modeling two-body scattering of stars in galactic nuclei (typically $\sim10^{-4}$~yr$^{-1}$ per galaxy), which may in part be due to the fact that only a small fraction of TDEs are optically luminous \citep{stone14,metzger15}. Selection effects from optical surveys may also play a role in this discrepancy, and a careful determination of TDE detection efficiency and completeness is needed to determine the degree to which the actual TDE rate may be higher than the rates inferred from optical surveys.


\begin{figure}
\centering
\subfloat{{\includegraphics[width=0.98\linewidth]{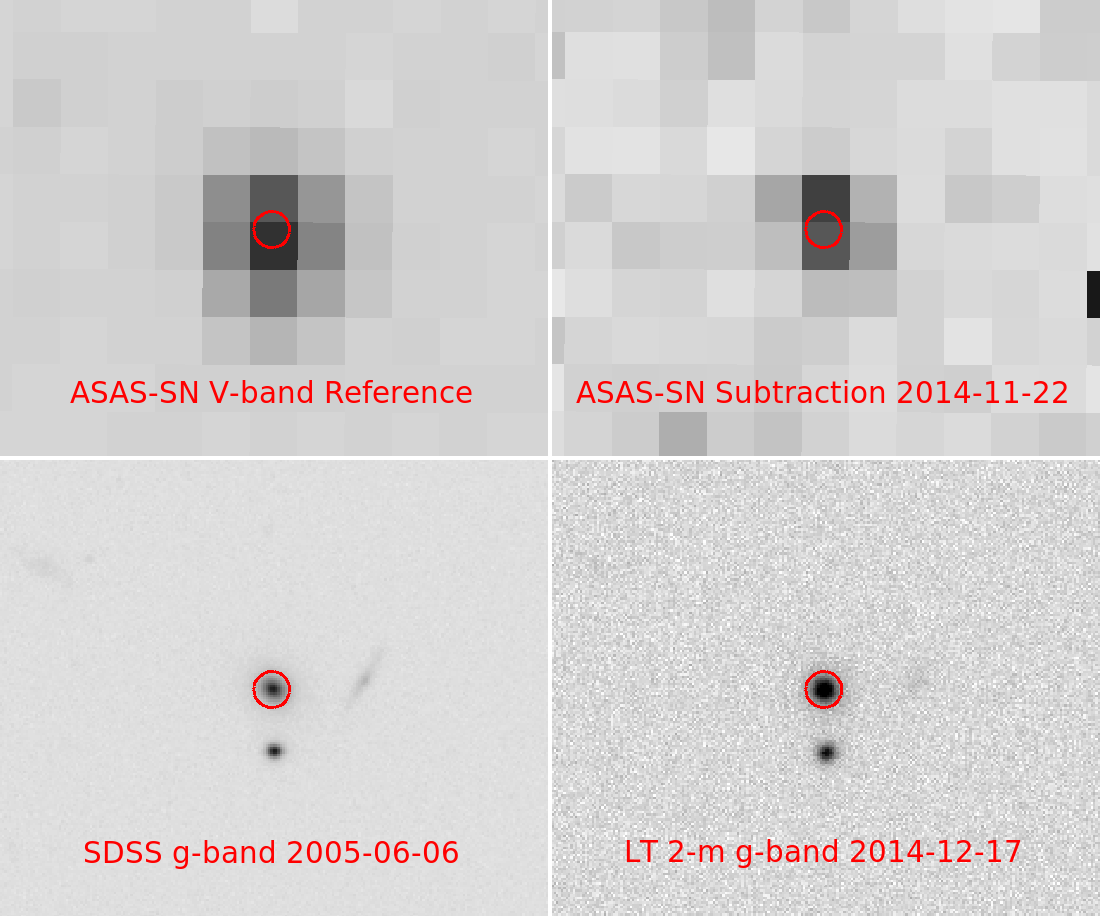}}}
\caption{Discovery images of ASASSN-14li. The top-left panel shows the ASAS-SN $V$-band reference image and the top-right panel shows the ASAS-SN subtracted image from 2014 November 22, the date of discovery. The bottom-left panel shows the archival SDSS $g$-band image of the host galaxy {\galname} and the right panel shows an LT 2-m $g$-band image taken during \edit{our follow-up campaign}. The dates of the observations are listed on each panel. The red circles have radii of 3\farcs{0} and are centered on the position of the transient measured in the LT image.}
\label{fig:finding_chart}
\end{figure}

In this manuscript we describe the discovery and follow-up observations of {\name}, a candidate TDE discovered by the All-Sky Automated Survey for SuperNovae (ASAS-SN\footnote{\url{http://www.astronomy.ohio-state.edu/~assassin/}}; \citealt{shappee14}). ASAS-SN is a long-term project to monitor the whole sky on a rapid cadence to find nearby supernovae \citep[e.g.,][]{holoien15a} and other bright transients, such as AGN activity \citep[e.g.,][]{shappee14}, extreme stellar flares \citep[e.g.,][]{schmidt14}, outbursts in young stellar objects \citep[e.g.,][]{holoien14a}, and cataclysmic variable stars \citep[e.g.,][]{kato13,kato14}. Our transient source detection pipeline was triggered on 2014 November 22, detecting a new source with $V=16.5\pm0.1$~mag \citep{asassn14li_atel}. The object was also detected on 2014 November 11 at  $V=15.8\pm0.1$~mag, but is not detected ($V\ga17$~mag) in data obtained on 2014 July 13 and before. Unfortunately, no data were obtained between 2014 July 13 and 2014 November 11 as the galaxy was behind the Sun. 

A search at the object's position (J2000 RA/Dec $=$ 12:48:15.23/+17:46:26.22) in the Sloan Digital Sky Survey Data Release 9 \citep[SDSS DR9;][]{ahn12} catalog revealed the source of the outburst to be the galaxy {\galname} (VII Zw 211) at redshift $z=0.0206$, corresponding to a luminosity distance of $d=90.3$~Mpc ($H_0=73$~km~s$^{-1}$~Mpc$^{-1}$, $\Omega_M=0.27$, $\Omega_{\Lambda}=0.73$), and that the ASAS-SN source position was consistent with the center of the host galaxy. Follow-up images obtained on 2014 November 28 with the Las Cumbres Observatory Global Telescope Network (LCOGT) 1-m telescope at McDonald Observatory \citep{brown13} and on 2014 November 30 with the {\swift} UltraViolet and Optical Telescope \citep[UVOT;][]{roming05} confirmed the detection of the transient.

In order to constrain any offset between the source of the outburst and the nucleus of the host galaxy we first astrometrically aligned an image of the transient taken with the 2-m Liverpool Telescope \citep[LT;][]{steele04} with the archival SDSS image of the host galaxy. From this aligned image, we measure an offset of $0.43\pm0.52$~pixels ($0.17\pm0.21$~arcseconds, or $74.4\pm91.9$~parsecs) between the position of the brightest pixel in the host galaxy in the LT image and the position of the brightest pixel in the SDSS image. This offset is consistent with the source of the outburst being the nucleus of the host galaxy, which provides support for a TDE interpretation of the event. Figure~\ref{fig:finding_chart} shows the ASAS-SN $V$-band reference image of the host galaxy and the ASAS-SN $V$-band subtraction image from the discovery epoch as well as archival SDSS and post-discovery LT $g$-band images.

The archival SDSS spectrum of {\galname} shows little evidence of strong AGN activity. A follow-up spectrum of the nuclear region of the host obtained on 2014 November 30 with the SuperNova Integral Field Spectrograph \citep[SNIFS;][]{lantz04} mounted on the University of Hawaii 2.2-m telescope showed a broad H$\alpha$ emission feature at the redshift of the host with a $\rm FWHM \simeq9000$~km~s$^{-1}$ and increased emission at bluer wavelengths. In addition to this follow-up spectrum, follow-up photometry of the source obtained on 2014 November 30 with the {\swift} X-ray Telescope \citep[XRT;][]{burrows05} and UVOT showed strong soft X-ray emission and ultraviolet emission from a location consistent with the host nucleus. Given these observations, we determined that {\name} was a potential tidal disruption event, and began an extensive follow-up campaign in order to characterize the transient.

In \S\ref{sec:obs} we describe pre-outburst data, including photometry and spectroscopy, of the host galaxy as well as new observations obtained of the transient during our follow-up campaign. In \S\ref{sec:analysis} we analyze these data to model the transient's luminosity and temperature evolution and compare the properties of {\name} to those of other TDE candidates in literature. Finally, in \S\ref{sec:rates}, we use the TDE discoveries by ASAS-SN to estimate the rate of these transients in the nearby universe.


\section{Observations and Survey Data}
\label{sec:obs}

In this section we summarize the available archival survey data of the transient host galaxy {\galname} as well as our new photometric and spectroscopic observations of {\name}.


\subsection{Archival Data}
\label{sec:arch_dat}

We retrieved archival reduced $ugriz$ images of {\galname} from SDSS DR9 and measured the fluxes in a 5\farcs{0} aperture radius. This aperture radius was also used to measure the source flux in follow-up data, and was chosen to match the {\swift} PSF. We also obtained near-IR $JHK_s$ images from the Two-Micron All Sky Survey \citep[2MASS;][]{skrutskie06} and measured 5\farcs{0} aperture magnitudes using the same procedure. Finally, we obtained a near-UV image of the host galaxy from the Galaxy Evolution Explorer (GALEX) Data Release 7 and measured the magnitude of the host using PSF photometry. The measured fluxes from GALEX, SDSS, and 2MASS were then used for host galaxy SED modeling and for subtracting the host galaxy flux from follow-up data containing the transient. We present the measured 5\farcs{0} aperture magnitudes from the SDSS and 2MASS images and the PSF magnitude from the GALEX image in Table~\ref{table:host_mags}.


\begin{table}
\centering
\caption{Photometry of the Host Galaxy}
\label{table:host_mags}
\begin{tabular}{@{}ccc}
\hline
Filter & Magnitude & Magnitude Uncertainty \\
\hline
$NUV$ & 19.77 & 0.06 \\
$u$ & 17.61 & 0.03 \\
$g$ & 16.15 & 0.03 \\
$r$ & 15.63 & 0.03 \\
$i$ & 15.37 & 0.03\\
$z$ & 15.17 & 0.03 \\
$J$ & 14.30 & 0.07 \\
$H$ & 13.48 & 0.07 \\
$K_s$ & 13.20 & 0.09 \\
\hline
\end{tabular}

\medskip
\raggedright
\noindent 
These are 5\farcs{0} radius aperture magnitudes from GALEX, SDSS, and 2MASS.
\end{table}

\begin{figure*}
\begin{minipage}{\textwidth}
\centering
\subfloat{{\includegraphics[width=0.48\textwidth]{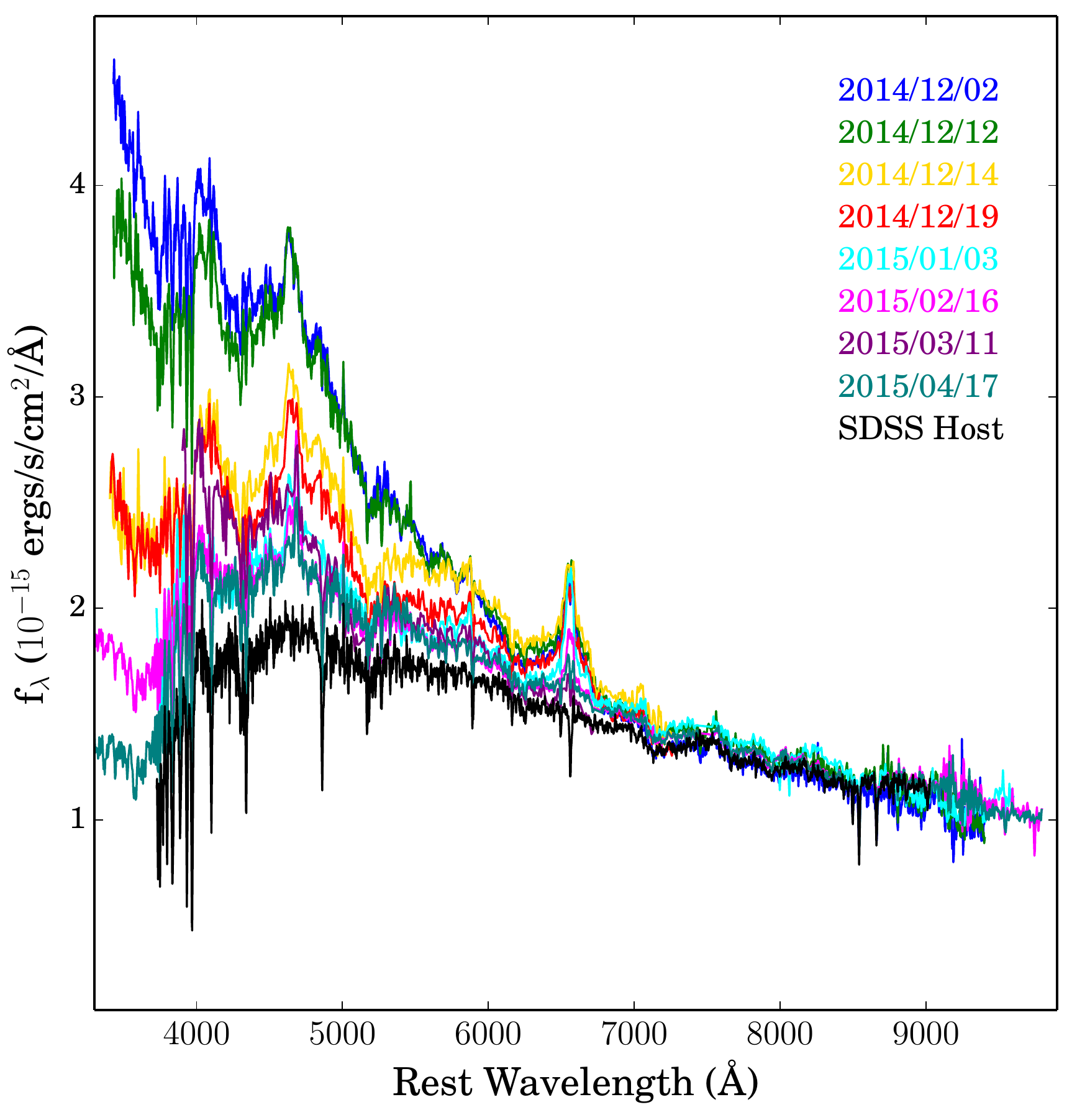}}}
\subfloat{{\includegraphics[width=0.48\textwidth]{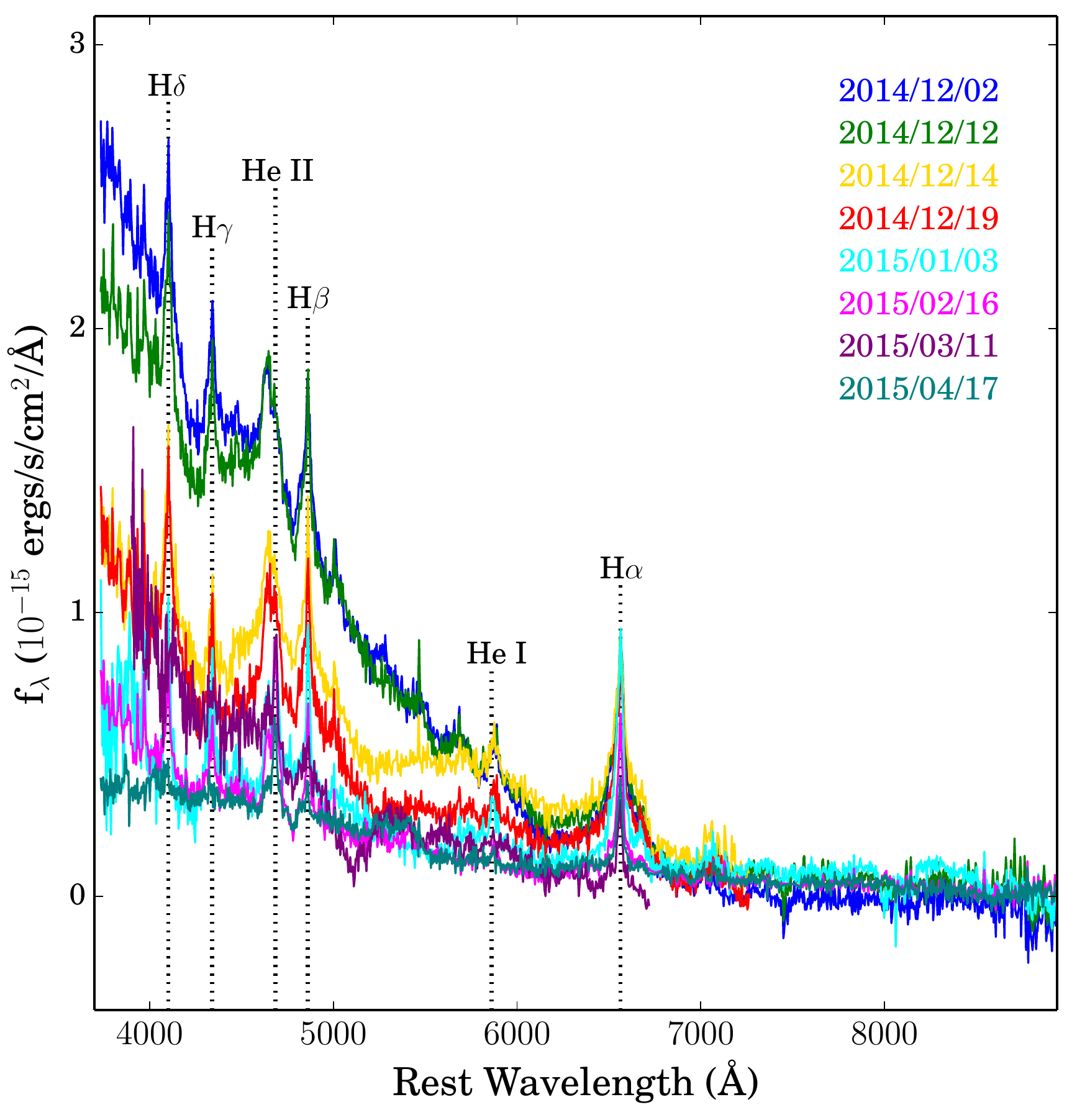}}}
\caption{\emph{Left Panel}: Spectral time-sequence for {\name}. Colors indicate the dates of observation. The archival SDSS spectrum of {\galname} is plotted in black. \emph{Right Panel}: Host-galaxy-subtracted spectral time-sequence of {\name}, showing the same spectra used in the left panel with the host galaxy spectrum subtracted. Prominent emission lines in the transient spectrum, including Balmer lines, \ion{He}{1}~5725{\AA}, and \ion{He}{2}~4686{\AA}, are indicated by vertical dotted lines. The transient spectra show many broad emission features and blue continuum emission above that of the host at wavelengths shorter than $\sim 4000$~{\AA} in all epochs.}
\label{fig:spectra}
\end{minipage}
\end{figure*}

There are no archival Spitzer, Herschel, Hubble Space Telescope (HST), Chandra, or X-ray Multi-Mirror Mission (XMM-Newton) observations of {\galname}. Examining data from the ROSAT All-Sky Survey, we do not detect the host galaxy with a 3-sigma upper limit of $1.8\times10^{-2}$ counts~s$^{-1}$, corresponding to $7.5\times10^{-14}$~ergs~s$^{-1}$~cm$^{-2}$, in the $0.08-2.9$~keV band \citep{voges99}, which provides evidence that the galaxy does not host a strong AGN. However, the host galaxy is detected in the Very Large Array Faint Images of the Radio Sky at Twenty-cm survey \citep[FIRST;][]{becker95}, with 1.4 GHz flux density of $2.96\pm0.15$ mJy, corresponding to a luminosity of $L_{1.4 GHz}\sim2.6\times10^{21}$~W~Hz$^{-1}$. If this radio emission were caused by star formation, the far-IR-radio correlation would imply a FIR luminosity of $L_{FIR}\sim3\times10^9$~{\lsun} \citep{yun01}. However, detections of the host in archival Wide-field Infrared Survey Explorer \citep[WISE:][]{wright10} data gives a $W3$-band magnitude of $m_{W3}=12.367\pm0.439$, corresponding to a luminosity of $L_{W3}\sim2\times10^7$~{\lsun}. This implies that the FIR luminosity would need to be roughly two orders of magnitude greater than the mid-IR (MIR) luminosity in order to reach the expected value if the radio emission was caused by star formation. In addition, if the galaxy exhibited the strong star formation implied by this radio emission, we would expect to see additional signs of star formation, such as strong [\ion{O}{2}]~3727\AA~emission, which we do not see in the host spectrum (see Figure~\ref{fig:spectra}). Thus, the radio emission from the host is likely not related to star formation, but may instead be an indicator of nuclear activity. However, the galaxy has a mid-IR (MIR) color of $(W1-W2)\simeq0.01\pm0.04$ in the WISE data, which, along with the non-detection in the X-ray data, provides evidence that any AGN activity is not strong \citep[e.g.,][]{assef13}.


\begin{figure*}
\begin{minipage}{\textwidth}
\centering
\subfloat{{\includegraphics[width=0.7\textwidth]{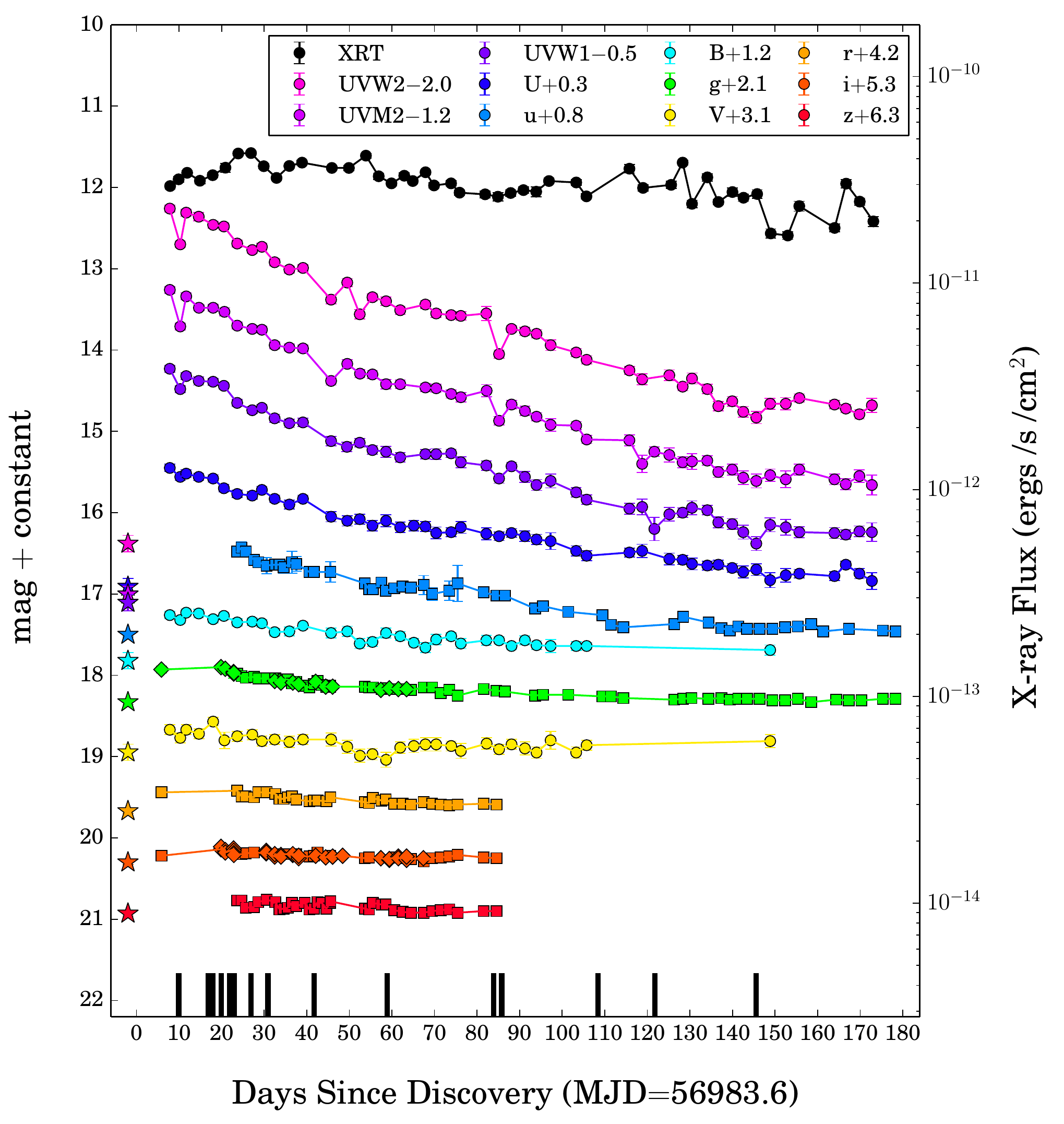}}}
\caption{Light curves of {\name} beginning on the epoch of discovery (MJD$=56983.6$) and spanning 178 days. Follow-up data obtained from {\swift} (X-ray, UV, and optical; \edit{circles}), the LT 2-m (optical; \edit{squares}), and the LCOGT 1-m telescopes (optical; \edit{diamonds}) are shown as circles. All UV and optical magnitudes are shown in the Vega system (left scale), and X-ray fluxes are shown in ergs/s/cm$^2$ (right scale). The scales are chosen so that time variability has the same meaning for both the X-ray and optical/UV data. The data are not corrected for extinction and error bars are shown for all points, but in some cases they are smaller than the data points. Observations in the $riz$ and {\swift} $BV$ filters were terminated earlier than those in other bands, as the source had faded to be fainter than the host. 5\farcs0 aperture magnitudes measured from archival SDSS images for $ugriz$ and synthesized from our host SED model for the {\swift} UVOT filters are shown as stars at $-5$ days. Vertical bars at the bottom of the figure indicate dates of spectroscopic follow-up. Although the peak of the light curve was likely missed due to the source being in an unobservable position behind the Sun, the data still show that {\name} brightened considerably in the UV and in the bluer optical filters, with the largest increase being slightly more than 4 magnitudes in the {\swift} $UVW2$ band. Table~\ref{tab:phot} contains all the follow-up photometric data.}
\label{fig:lightcurve}
\end{minipage}
\end{figure*}

Using the code for Fitting and Assessment of Synthetic Templates \citep[FAST v1.0;][]{kriek09}, we fit stellar population synthesis (SPS) models to the 5\farcs{0} SDSS $ugriz$ and 2MASS $JHK_s$ magnitudes of the host galaxy. The fit was made assuming a  \citet{cardelli88} extinction law with $R_V=3.1$ and Galactic extinction of $A_V=0.07$~mag based on \citet{schlafly11}, an exponentially declining star-formation history, a Salpeter IMF, and the \citet{bruzual03} models. We obtained a good SPS fit (reduced $\chi_{\nu}^{2}=0.6$), with the following parameters: $M_{*}=(2.8_{-0.1}^{+0.1})\times10^{9}$~{\msun}, age$=1.3_{-0.3}^{+0.2}$~Gyr, and a $1\sigma$ upper limit on the star formation rate of $\textrm{SFR}\leq0.9\times 10^{-2}$~{\msun}~yr$^{-1}$. In order to estimate the mass of the SMBH in {\galname}, we use a bulge mass of $M_B\sim10^{9.3}$~{\msun} based on the total mass from the FAST fit and \citet{mendel14} and the $M_{B}$-$M_{BH}$ relation from \citet{mcconnell13}, giving $M_{BH}\sim10^{6.7}$~{\msun}, a value very similar to that estimated for the host of ASASSN-14ae \citep{holoien14b}. We find no evidence for any significant additional extinction in fits to the transient spectral energy distribution (SED) despite the fact that the {\swift} UV data, particularly the $UVM2$ band which lies on top of the 2200~\AA~extinction curve feature, is a powerful probe for additional dust. In the analyses event's SED which follow, we only correct for Galactic extinction.

We also retrieved an archival spectrum of {\galname} from SDSS DR9. The archival spectrum shows both [\ion{O}{3}]~5007\AA~and [\ion{N}{2}]~6584\AA~emission with luminosities of $L{[\ion{O}{3}]}\sim4.4\times10^5$~{\lsun} and $L{[\ion{N}{2}]}\sim3.6\times10^5$~{\lsun}, respectively. As with the radio emission, this emission is likely not related to star formation, as the host shows no H$\alpha$ or [\ion{O}{2}] in emission. Rather, this is again an indication that the host galaxy may host a low-luminosity AGN. However, the lack of X-ray emission and the MIR colors of the host from WISE imply that this nuclear activity is weak, as previously discussed.

The archival spectrum also shows H$\delta$ absorption with a large equivalent width, indicating a change in the galaxy's star formation history within the last $\sim$1 Gyr \citep[e.g.][]{goto03}. The strong H$\delta$ absorption indicates that the SED of the host is dominated by A stars, which implies that {\galname} may be a post-starburst galaxy \citep{goto03}. \edit{While some TDE hosts, such as NGC 5905, have shown evidence of nuclear star formation \citep{komossa99}, recent results from \citet{arcavi14} suggested that TDE candidates prefer post-starburst hosts, and {\galname} may follow this pattern. The relation between starbursts and TDEs will be further explored in a future paper on TDE hosts (Dong et al, \emph{in prep.}).}


\subsection{New Photometric Observations}
\label{sec:phot}

Following the discovery of the transient, we requested and were granted a series of 53 {\swift} XRT and UVOT target-of-opportunity (ToO) observations \edit{between 2014 November 30 and 2015 May 14}. The UVOT observations were obtained in 6 filters: $V$ (5468~\AA), $B$ (4392~\AA), $U$ (3465~\AA), $UVW1$ (2600~\AA), $UVM2$ (2246~\AA), and $UVW2$ (1928~\AA) \citep{poole08}. We extracted source counts from a 5\farcs0 radius region and sky counts from a $\sim$40\farcs0 radius region using the UVOT software task {\sc uvotsource}. We then used the most recent UVOT calibrations \citep{poole08,breeveld10} to convert these count rates into magnitudes and fluxes. 


The XRT operated in Photon Counting mode \citep{hill04} during our observations. The data from all epochs were reduced with the software tasks {\sc xrtpipeline} and {\sc xrtselect}. We extracted X-ray source counts and background counts using a region with a radius of 20 pixels (47\farcs{1}) centered on the source position and a source-free region with a radius of 100 pixels (235\farcs{7}), respectively. In all epochs of observation, we detect X-ray emission consistent with the position of the transient. To convert the detected counts to fluxes, we assume a power law spectrum with $\Gamma=2$ and Galactic \ion{H}{1} column density \citep{kalberla05}.

\edit{The XRT data have an average detected count rate of 0.3 counts~s$^{-1}$ in the 0.3$-$10 keV range, corresponding to a flux of $3.0\times10^{-11}$~ergs~s$^{-1}$~cm$^{-2}$. This is roughly equivalent to a count rate of 1.2 counts~s$^{-1}$ in the ROSAT PSPC, indicating an increase by a factor of $\sim60$ over the ROSAT limits.}

In addition to the {\swift} XRT and UVOT observations, we also obtained $ugriz$ images with the LT 2-m telescope and the LCOGT 1-m telescopes at Siding Spring Observatory, McDonald Observatory, Sutherland, and Cerro Tololo. We measured aperture photometry using a 5\farcs0 aperture radius in order to match the host galaxy and {\swift} measurements\footnote{We attempted to do image subtraction of the follow-up $ugriz$ data with the SDSS archival images as templates. However, due to the lack of stars in the field-of-view close to {\name}, the quality of the subtractions was sub-optimal and accurate measurements were not possible.}. We determined photometric zero-points using several SDSS stars in the field. 

Figure~\ref{fig:lightcurve} shows the X-ray, UV, and optical light curves of {\name}. The XRT flux measurements and UVOT/$ugriz$ magnitudes are presented in Table~\ref{tab:xray} and Table~\ref{tab:phot}, respectively. The observations cover the period from MJD 56983.6 (the epoch of discovery) through our latest epoch of observations on MJD 57161.9, spanning 178.3 days. The data are shown without extinction correction or host flux subtraction. Also shown in Figure~\ref{fig:lightcurve} are the host magnitudes measured from SDSS images for $ugriz$ and extrapolated from the host SED fit for {\swift} UVOT filters. Although our observations likely missed the peak of the transient's light curve, they still show that {\name} brightened considerably with respect to the host galaxy in the UV and blue filters, with the largest increase being in the {\swift} $UVW2$ band, where it brightened by $\Delta m_{UVW2}\sim-4.1$. The $g$-band increase was significantly weaker, with $\Delta m_{g}\sim-0.4$. 


\subsection{New Spectroscopic Observations}
\label{sec:spec}

We obtained spectra of {\name} spanning 145 days between UT 2014 December 02 and UT 2015 April 14. The spectrographs used for these observations, \edit{along with wavelength range and resolution in angstroms,} were: SNIFS mounted on the 2.2-m University of Hawaii telescope ($3200-10000$~\AA, $\rm R\sim 3$~\AA), the Dual Imaging Spectrograph (DIS) mounted on the Apache Point Observatory 3.5-m telescope (range $3500-9800$~\AA, $\rm R\sim 7$~\AA), the Multi-Object Double Spectrographs (MODS; \citealt{Pogge2010}) mounted on the dual 8.4-m Large Binocular Telescope (LBT) on Mount Graham ($3200-10000$~\AA, $\rm R\sim 3$~\AA), the Modular Spectrograph (Modspec) mounted on the MDM Observatory Hiltner 2.4-m telescope ($4660-6730$~\AA, $\rm R\sim 4$~\AA), the Ohio State Multi-Object Spectrograph \citep[OSMOS;][]{martini11} mounted on the MDM Observatory Hiltner 2.4-m telescope ($4200-6800$~\AA, $\rm R\sim 4$~\AA), the Fast Spectrograph \citep[FAST;][]{fabricant98} mounted on the Fred L. Whipple Observatory Tillinghast 1.5-m telescope ($3700-9000$~\AA, $\rm R\sim 3$~\AA), and the Inamori Magellan Areal Camera and Spectrograph \citep[IMACS;][]{dressler11} mounted on the Las Campanas Observatory Magellan-Baade 6.5-m telescope ($3650-9740$~\AA, $\rm R\sim 8$~\AA). The spectra from MODS were reduced using a custom pipeline written in IDL\footnote{\url{http://www.astronomy.ohio-state.edu/MODS/Software/modsIDL/}} while all other spectra were reduced using standard techniques in IRAF. Telluric corrections were applied to the spectra using the spectra of spectrophotometric standard stars observed on the same nights. We calculated synthetic $r$-band magnitudes and scaled each spectrum to match the $r$-band photometry. Figure~\ref{fig:spectra} shows a time-sequence of \edit{selected} flux-calibrated follow-up spectra along with the archival SDSS spectrum of the host as well as a time-sequence of the same spectra with the host galaxy spectrum subtracted. Summary information, including dates, instruments used, and exposure times for \edit{all follow-up spectra}, are listed in Table~\ref{tab:spectra}.

The key characteristics of the spectra of {\name} are a strong blue continuum, consistent with the photometric measurements, and the presence of broad Balmer and helium lines in emission, which are either absent or seen as absorption features in the host spectrum. The emission features are highly asymmetric and have widths of $\sim10,000-20,000$~km~s$^{-1}$ in all epochs, though they appear to grow narrower over time. The blue continuum becomes progressively weaker over time, though it still shows emission in excess of the host at wavelengths shorter than $\sim6000$~{\AA} in the latest epoch, which is in agreement with the UV and optical photometry. We further analyze the features of these spectra and compare to ASASSN-14ae and other TDE candidates in \S\ref{sec:specanal}.


\section{Analysis}
\label{sec:analysis}


\subsection{SED Analysis}
\label{sec:sedanal}

\begin{figure}
\centering
\subfloat{{\includegraphics[width=0.95\linewidth]{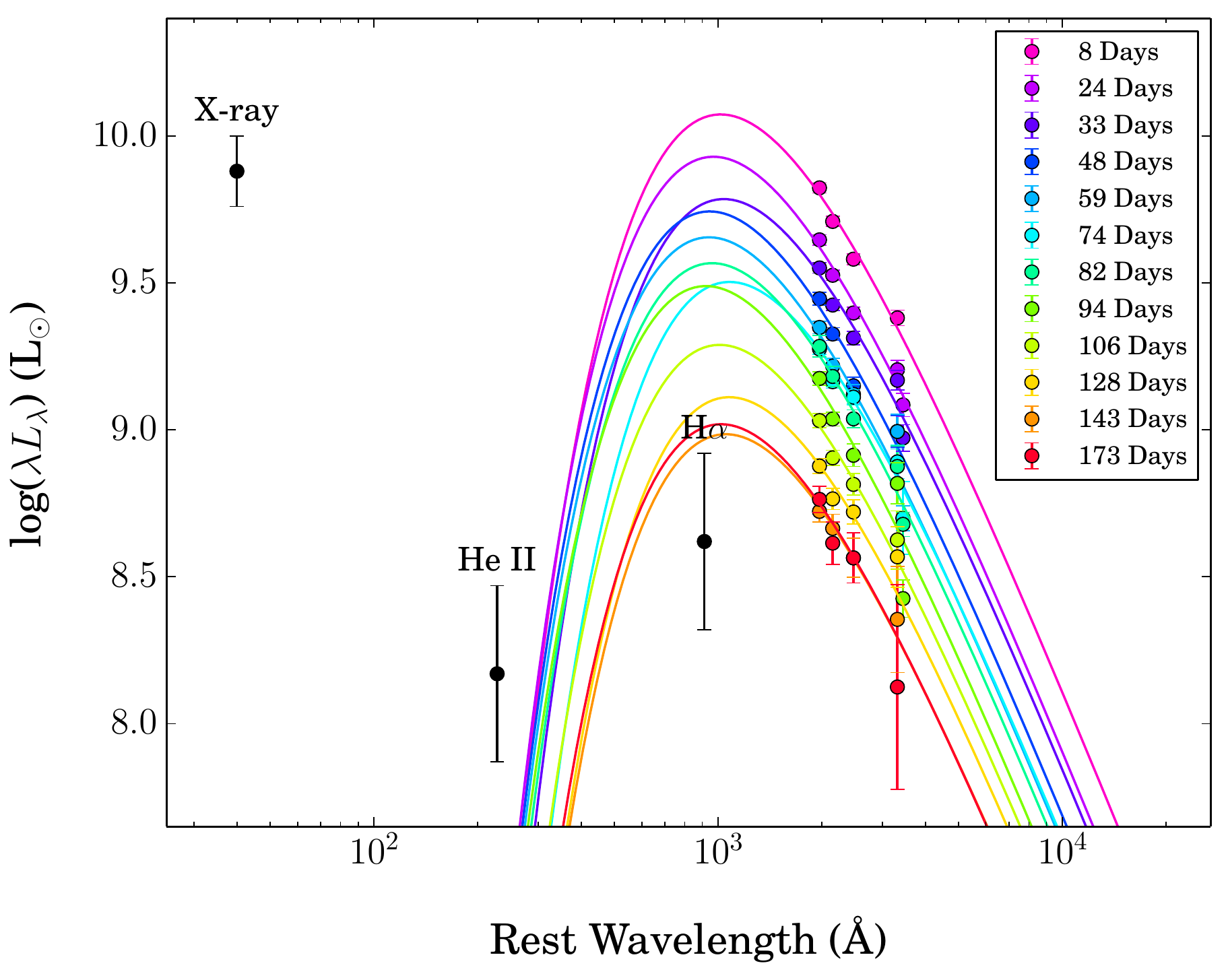}}}
\caption{Evolution of the SED of {\name} shown in different colors (points) and the corresponding best-fitting blackbody models for each epoch (lines). \edit{Only data points with subtracted flux greater than 0.3 times the host flux are shown, as data points with a smaller fractional flux were highly uncertain and contributed little to the fit.} All data points have been corrected for Galactic extinction and include error bars, though in some cases the error bars can be smaller than the data points. All fits were made assuming a temperature prior of $\log{T/K}=4.55\pm0.05$. Also shown are the early-epoch X-ray luminosity and ionizing luminosities implied by the H$\alpha$ and \ion{He}{2} 4868{\AA} lines (see \S~\ref{sec:specanal}.) No additional UV emission is required to produce the line emission, but the X-ray emission is likely non-thermal or produced in a separate, hotter region.}
\label{fig:sed_evol}
\end{figure}

Using the 5\farcs0 aperture magnitudes measured from the archival SDSS data and synthesized from the FAST SED fit for {\swift} UVOT filters we performed host flux subtraction on the extinction-corrected photometric follow-up data. We then used these host-subtracted fluxes to fit the SED of {\name} with blackbody curves using Markov Chain Monte Carlo (MCMC) methods, as was done for ASASSN-14ae in \citet{holoien14b}. As flux in redder filters was clearly dominated by host flux even in early epochs, we only use filters with effective wavelength less than 4000 {\AA} ({\swift} $U$, $UVW1$, $UVM2$, and $UVW2$ and SDSS $u$) when fitting the SED. Unlike ASASSN-14ae, the UV data for {\name} do not appear to span the peak of the SED, resulting in a broad range of possible blackbody temperatures. If we use a very weak temperature prior ($\pm 1$~dex), the median epoch with measurements at four or more wavelengths has a formal temperature uncertainty of $\pm 0.2$~dex.  However, without a clear detection of the spectral peak, temperature uncertainties are likely dominated by systematic errors (e.g. host flux or deviations of the true spectral shape from a black body) rather than this estimate of the statistical errors. We therefore adopted a temperature typical of these weakly constrained fits and a strong prior of $\log{T/K}=4.55\pm0.05$ for our standard fits. The evolution of the source's SED along with the best-fit blackbody curves are shown in Figure~\ref{fig:sed_evol}. 


\begin{figure}
\centering
\subfloat{{\includegraphics[width=0.95\linewidth]{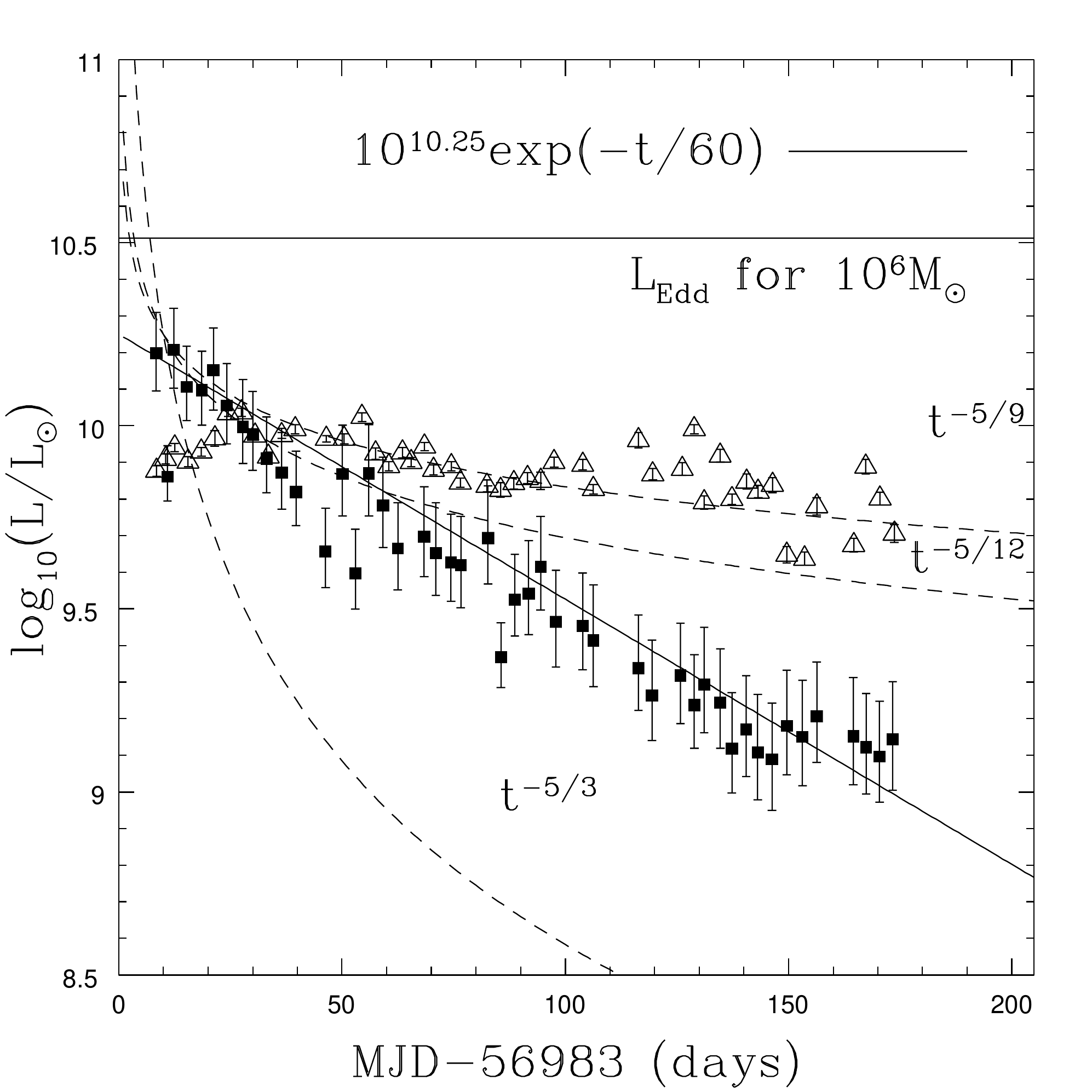}}}
\caption{Evolution of \name's luminosity over time. Open triangles indicate the X-ray luminosity measurements. Dashed lines show popular power law fits for TDE luminosity curves $L\propto t^{-x}$ \citep[e.g.,][]{strubbe09,lodato11} while the diagonal solid line shows an exponential fit. The Eddington luminosity for a $M=10^6$~{\msun} black hole is shown as a solid horizontal line. The luminosity evolution appears to be best-fit by the exponential model, similar to the previous ASAS-SN TDE candidate, ASASSN-14ae \citep{holoien14b}.}
\label{fig:lum_evol}
\end{figure}

With the temperature constrained by our prior to remain roughly constant, the optical/UV luminosity of the source fades steadily over the $\sim178$ days after initial discovery. The luminosity evolution is well fit by an exponential curve $L\propto e^{-t/t_0}$ with $t_0\simeq60$~days, as shown in Figure~\ref{fig:lum_evol}. This differs from common TDE models, where the luminosity evolution is expected to follow a power law $t^{-x}$ with $x\simeq5/12 - 5/3$ \citep[e.g.,][]{strubbe09,lodato11}. However, the exponential luminosity evolution matches that of ASASSN-14ae, the other ASAS-SN TDE \citep{holoien14b}, meaning that this behavior is consistent with previously discovered TDE candidates. This temperature and luminosity behavior is inconsistent with what would be expected if {\name} were a supernova (which typically show rapidly declining temperatures along with constant or declining luminosity \citep[e.g.,][]{miller09,botticella10,inserra13,graham14}, providing evidence that {\name} is a better match to a TDE than a supernova. 


\begin{figure}
\centering
\subfloat{{\includegraphics[width=0.95\linewidth]{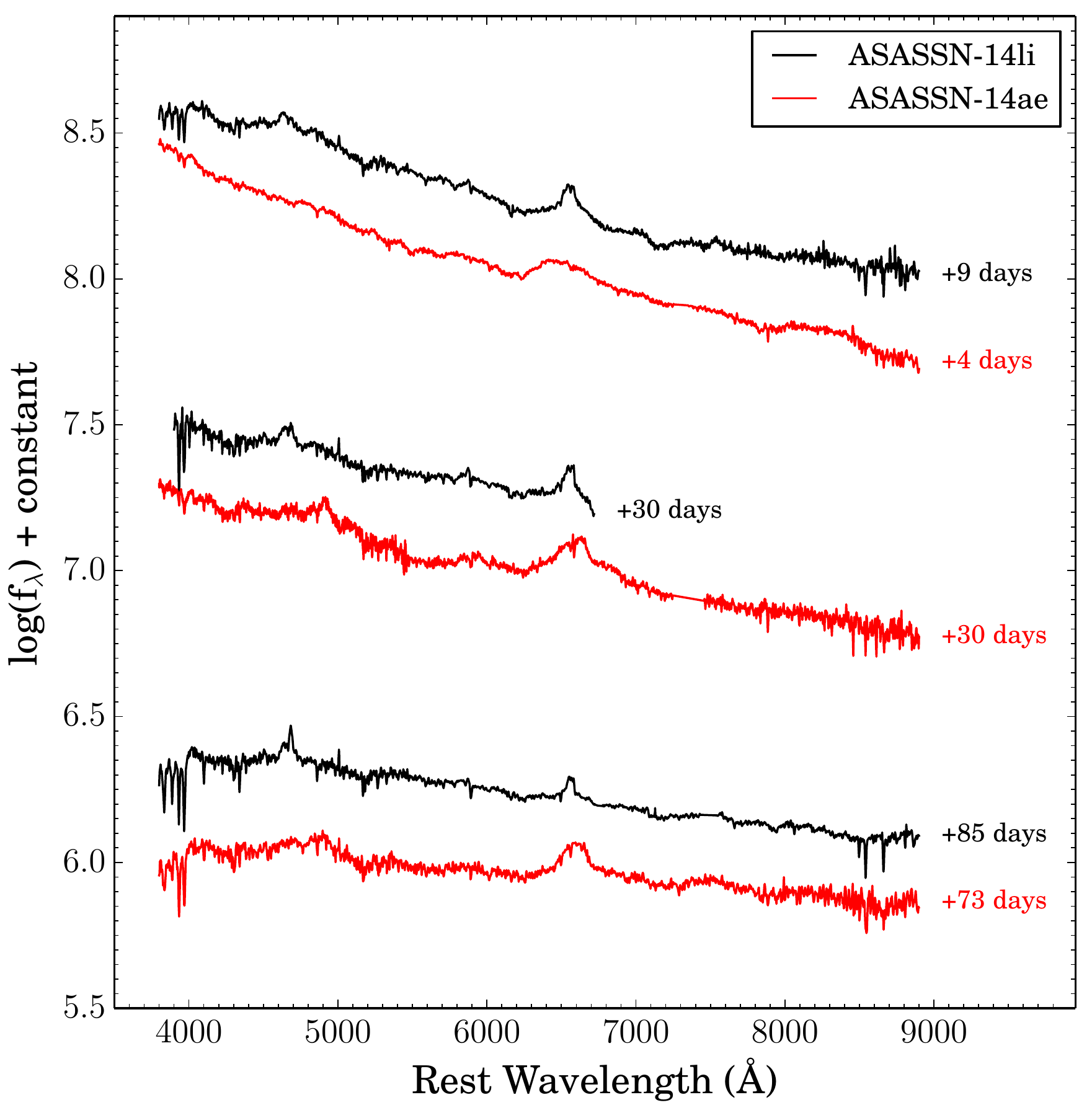}}}
\caption{Comparison of the spectra of {\name} with those of ASASSN-14ae \citep{holoien14b} at similar epochs after discovery. {\name} is shown in blue while ASASSN-14ae is shown in red, and epochs in days-after-discovery are shown to the right of each spectrum. The spectra look broadly similar, with both objects showing a strong blue continuum that declines over time and a broad H$\alpha$ emission feature in all epochs. However, the spectra of {\name} also show a strong \ion{He}{2} 4686~{\AA} emission feature in all epochs, whereas ASASSN-14ae only begins to show \ion{He}{2} emission in later epochs.}
\label{fig:spec_comp}
\end{figure}

The X-ray luminosity, by contrast, declines at a much slower rate than the optical/UV luminosity, and roughly 40 days after discovery the X-ray luminosity becomes the dominant source of emission. As shown in Figure~\ref{fig:sed_evol}, the X-ray luminosity requires a significantly higher blackbody temperature (roughly $T\sim10^5$~K) than the $T\sim35,000$~K temperature that best-fits the optical/UV data. While we cannot rule out a single blackbody component with much higher temperature generating both the optical/UV and the X-ray emission, we believe it is more likely that the the X-ray emission arises from a different, hotter region of the source seen through a region of lower-than-average density, as described in \citet{metzger15}, or that it is non-thermal.

Integrating over the X-ray and optical/UV luminosity curves using only epochs with {\swift} UV data implies that {\name} radiated a total energy of $E\simeq7\times10^{50}$~ergs over the time period covered by our follow-up data. This requires accretion of only $\Delta M \sim 4.0\times10^{-3}\eta_{0.1}^{-1}$~{\msun} of mass, where $\eta_{0.1}=0.1\eta$ is the radiative efficiency, to power the event.


\subsection{Spectroscopic Analysis}
\label{sec:specanal}

In their analysis of TDE candidates discovered by the Palomar Transient Factory, \citet{arcavi14} found that all their candidates showed similar spectroscopic characteristics, including a strong blue continuum, and that their candidates spanned a continuum from H-rich to He-rich spectroscopic features. The follow-up spectra of {\name} show many of the same characteristics as the TDEs that \citet{arcavi14} refer to as ``intermediate H+He events,'' because its spectra have strong He and H emission features in all epochs. In Figure~\ref{fig:spec_comp} we compare the spectra of {\name} to ASASSN-14ae from \citet{holoien14b} at similar epochs after discovery. The spectra are broadly similar, as expected, with both showing strong emission at bluer wavelengths and broad H$\alpha$ emission features, but there are some notable differences. The blue continuum seems to fade somewhat more slowly in {\name} than in ASASSN-14ae, and {\name} shows a broad \ion{He}{2} 4686~{\AA} emission feature even in early epochs (a few days after discovery), whereas ASASSN-14ae only began to show this feature a few months after discovery. We note that some of these differences could stem from the fact that {\name} was further past its peak luminosity when discovered than ASASSN-14ae was (the previous non-detection of ASASSN-14ae was roughly 3 weeks prior to discovery, whereas {\name} was not observed for nearly 3 months before being discovered). However, given the range of TDE candidate properties found by \citet{arcavi14}, it is likely that some of the differences observed between these objects are unrelated to their age.

\begin{figure}
\centering
\subfloat{{\includegraphics[width=0.95\linewidth]{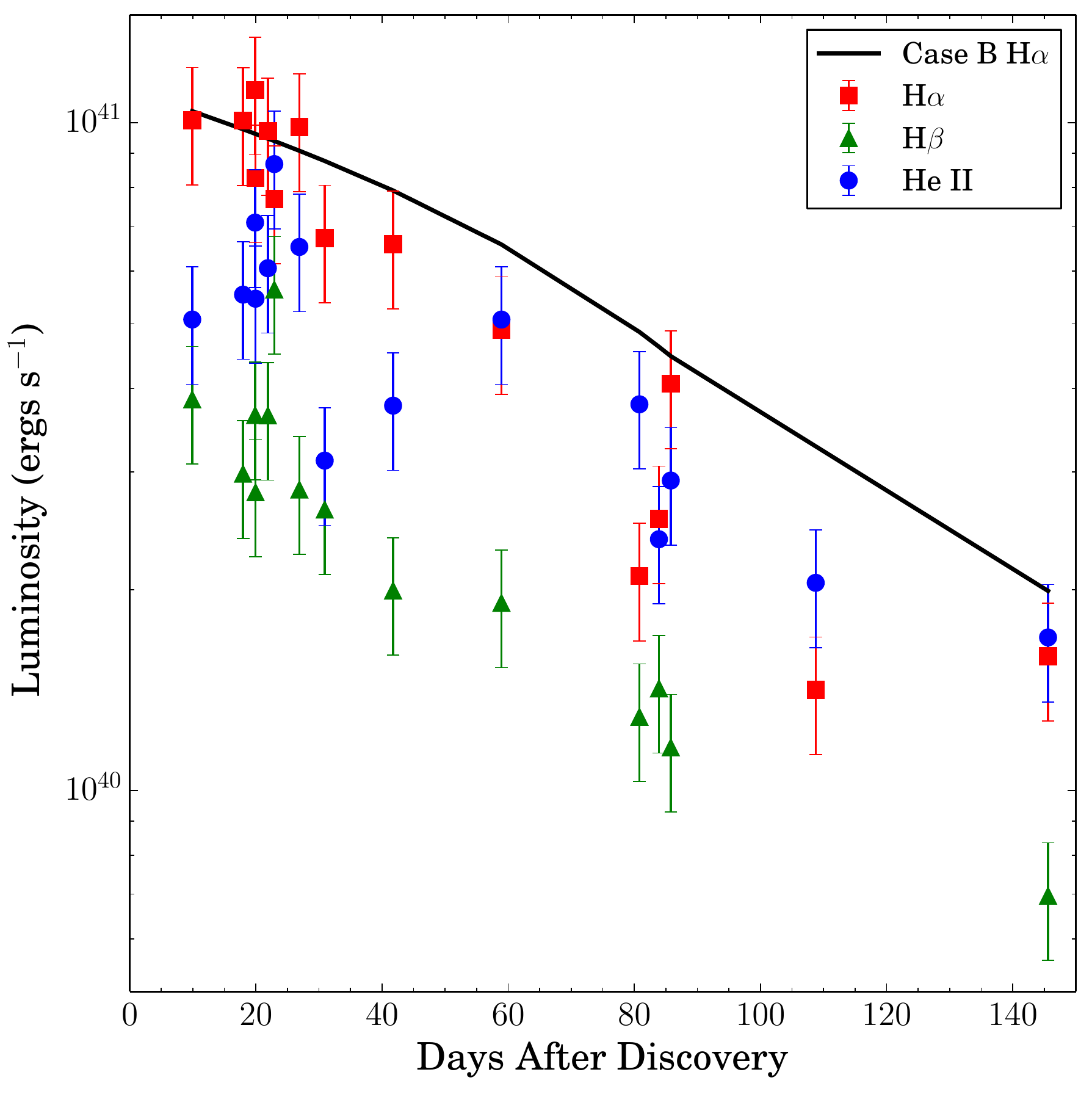}}}
\caption{Evolution of the H$\alpha$ (red squares), H$\beta$ (green triangles), and \ion{He}{2} 4686~{\AA} (blue circles) luminosities. Errorbars show 20\% errors on the line fluxes. The black line shows the H$\alpha$ emission that would be expected from case B recombination, given the H$\beta$ emission. The H$\alpha$ emission is generally consistent with recombination.}
\label{fig:line_lum}
\end{figure}

In Figure~\ref{fig:line_lum} we show the luminosity evolution of three strong emission features (H$\alpha$, H$\beta$, and \ion{He}{2} 4686~{\AA}) present in the spectra of {\name}. As estimating the true error on these fluxes is difficult given their complex shape, we assume 20\% errors on the emission fluxes calculated in each epoch. The three lines span a wide range in luminosity in early spectra, with peak values of $L_{H\alpha}\sim1.1\times10^{41}$~ergs~s$^{-1}$, $L_{H\beta}\sim4.0\times10^{40}$~ergs~s$^{-1}$, and $L_{\ion{He}{2}}\sim8.5\times10^{40}$~ergs~s$^{-1}$. However, the more luminous lines also appear to decline in luminosity more quickly, such that by the time of our most recent spectra, all three lines have luminosities in the range $L\sim(0.5-1.5)\times10^{40}$~ergs~s$^{-1}$. In Figure~\ref{fig:line_lum} we also show the H$\alpha$ emission expected given the measured H$\beta$ emission assuming the emission is driven by case B recombination. Within noise, the H$\alpha$/H$\beta$ ratio seems to be largely consistent with recombination. The measured luminosities for all three lines are given in Table~\ref{tab:line_lum}.


\begin{figure*}
\begin{minipage}{\textwidth}
\centering
\subfloat{{\includegraphics[width=0.48\textwidth]{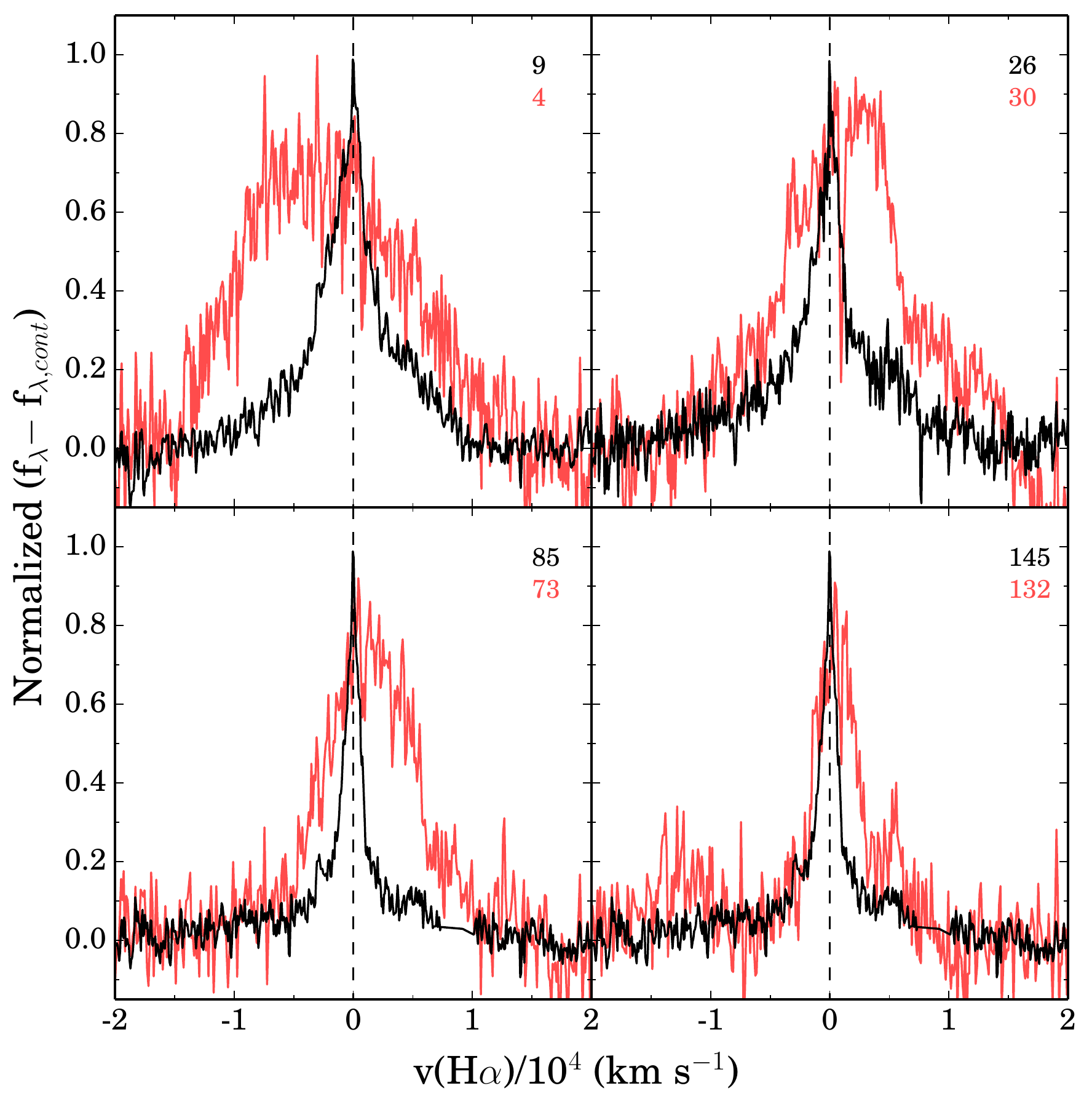}}}
\subfloat{{\includegraphics[width=0.48\textwidth]{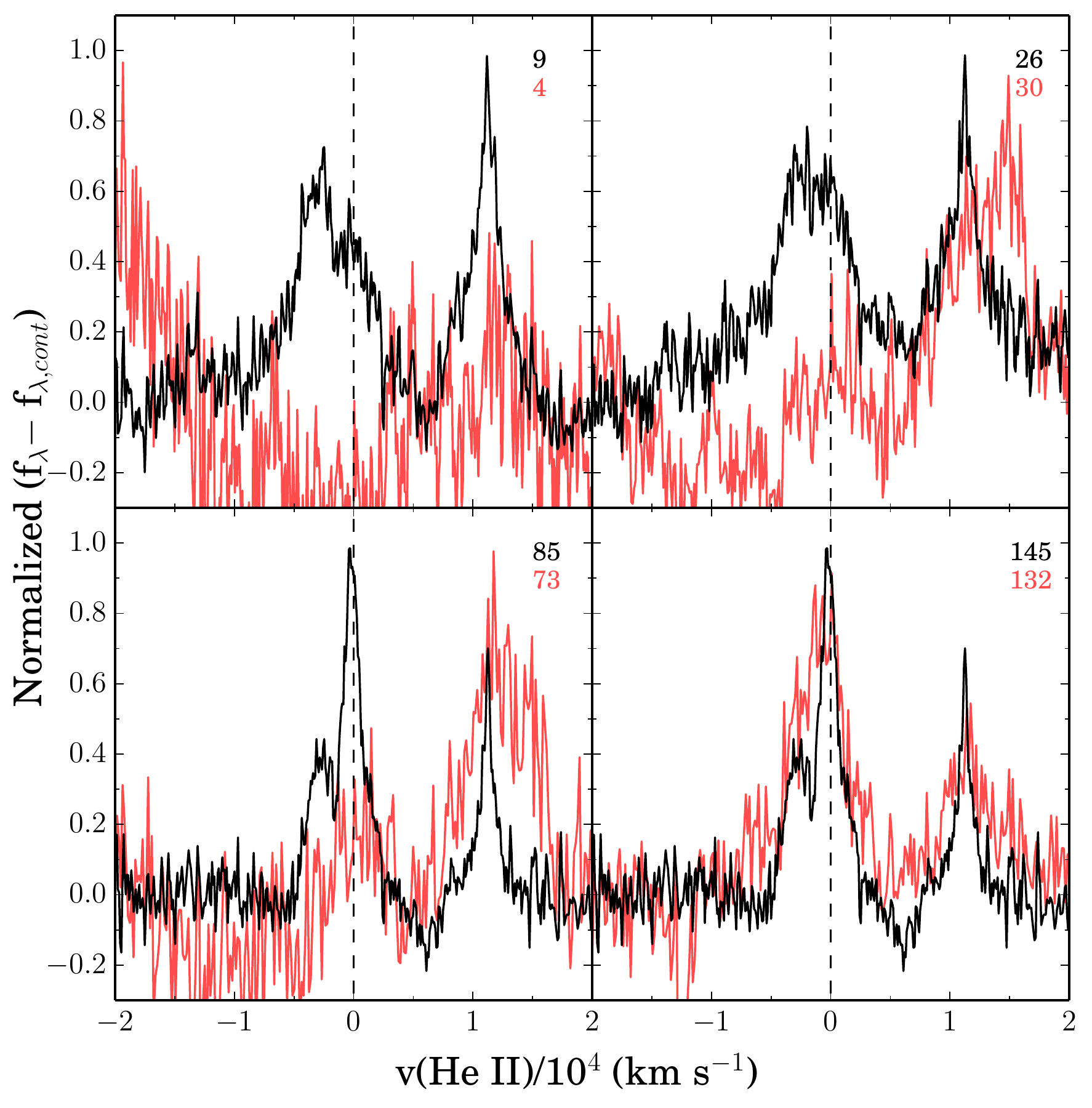}}}
\caption{Evolution of the H$\alpha$ (left panel) and \ion{He}{2} 4686~{\AA} (right panel) line profiles of {\name} (black) and ASASSN-14ae \citep[red;][]{holoien14b}. (The strong line to the right of the \ion{He}{2} 4686~{\AA} line in the right panel is H$\beta$.) The number of days since discovery for each spectrum is shown in the upper-right corner of each panel, with colors matching the colors of the spectra. We have subtracted the host galaxy spectra and a locally defined low-order continuum around the lines from each spectrum. Both objects show an asymmetric H$\alpha$ profile that narrows over time, with {\name} showing a significantly narrower profile than ASASSN-14ae in all epochs. {\name} shows a similar \ion{He}{2} profile that also narrows over time in all epochs, while ASASSN-14ae only shows significant \ion{He}{2} emission at later epochs. The \ion{He}{2} profile of {\name} shows two velocity peaks, one at $0$~km~s$^{-1}$ and one at $\sim-2000$~km~s$^{-1}$, which evolve in their relative intensity over time.}
\label{fig:line_comp}
\end{minipage}
\end{figure*}

We also compare the evolution of the H$\alpha$ and \ion{He}{2} 4686~{\AA} line profiles of {\name} to those of ASASSN-14ae in Figure~\ref{fig:line_comp}, spanning the period from 9 days after discovery to 145 days after discovery for {\name} and from 4 days after discovery to 132 days after discovery for ASASSN-14ae. While both objects show strong emission features, the evolution of these features shows a number of differences between the two objects. The H$\alpha$ profile is fairly asymmetric and narrows over time for both objects. However, {\name} shows a significantly narrower H$\alpha$ emission feature than ASASSN-14ae in all epochs. At 9 days, the {\name} H$\alpha$ feature shows a narrow peak and has blue/red wings reaching $\sim{-10,000}/{+10,000}$~km~s$^{-1}$ at the base of the line, while at 4 days the ASASSN-14ae H$\alpha$ feature shows a broad peak and has blue/red wings reaching $\sim{-15,000}/{+10,000}$~km~s$^{-1}$. The H$\alpha$ feature becomes significantly narrower for both objects by the time of the latest spectroscopic epoch, with {\name} showing blue/red wings reaching $\sim{-5,000}/{+5,000}$~km~s$^{-1}$ and ASASSN-14ae showing blue/red wings reaching $\sim{-5,000}/{+10,000}$~km~s$^{-1}$. The FWHM of the two lines show a similar evolutionary trend, with the H$\alpha$ feature narrowing from $\rm FWHM \simeq 3,000$~km~s$^{-1}$ to $\rm FWHM \simeq 1,500$~km~s$^{-1}$ for {\name} and from $\rm FWHM \simeq 17,000$~km~s$^{-1}$ to $\rm FWHM \simeq 8,000$~km~s$^{-1}$ for ASASSN-14ae. Given the fact that ``later'' epochs of ASASSN-14ae seem to resemble ``earlier'' epochs of {\name}, the H$\alpha$ evolution suggests that {\name} may have been older than ASASSN-14ae at discovery. However, the uncertainty on the age of {\name} is not so great as to suggest that all these differences are strictly due to the age of the transient.

The two TDE candidates show more obvious differences in the evolution of their \ion{He}{2} 4686~{\AA} line profiles. As can be seen in Figure~\ref{fig:line_comp}, {\name} displays a strong \ion{He}{2} emission feature in all epochs, while ASASSN-14ae only begins to show a similar feature in later spectra. The \ion{He}{2} feature shows similar evolution to the H$\alpha$ feature for {\name}, as it is fairly asymmetric in all epochs and narrows over time, with the 9-day spectrum showing blue/red wings reaching $\sim{-10,000}/{+5,000}$~km~s$^{-1}$ and the 145-day spectrum showing blue/red wings reaching $\sim{-5,000}/{+3,000}$~km~s$^{-1}$. ASASSN-14ae does not show a \ion{He}{2} feature in the 4-day spectrum, but its latest spectrum taken at 132 days after discovery shows a feature that resembles that of {\name}, with blue/red wings reaching $\sim{-5,000}/{+3,000}$~km~s$^{-1}$. 


\begin{figure}
\centering
\subfloat{{\includegraphics[width=0.95\linewidth]{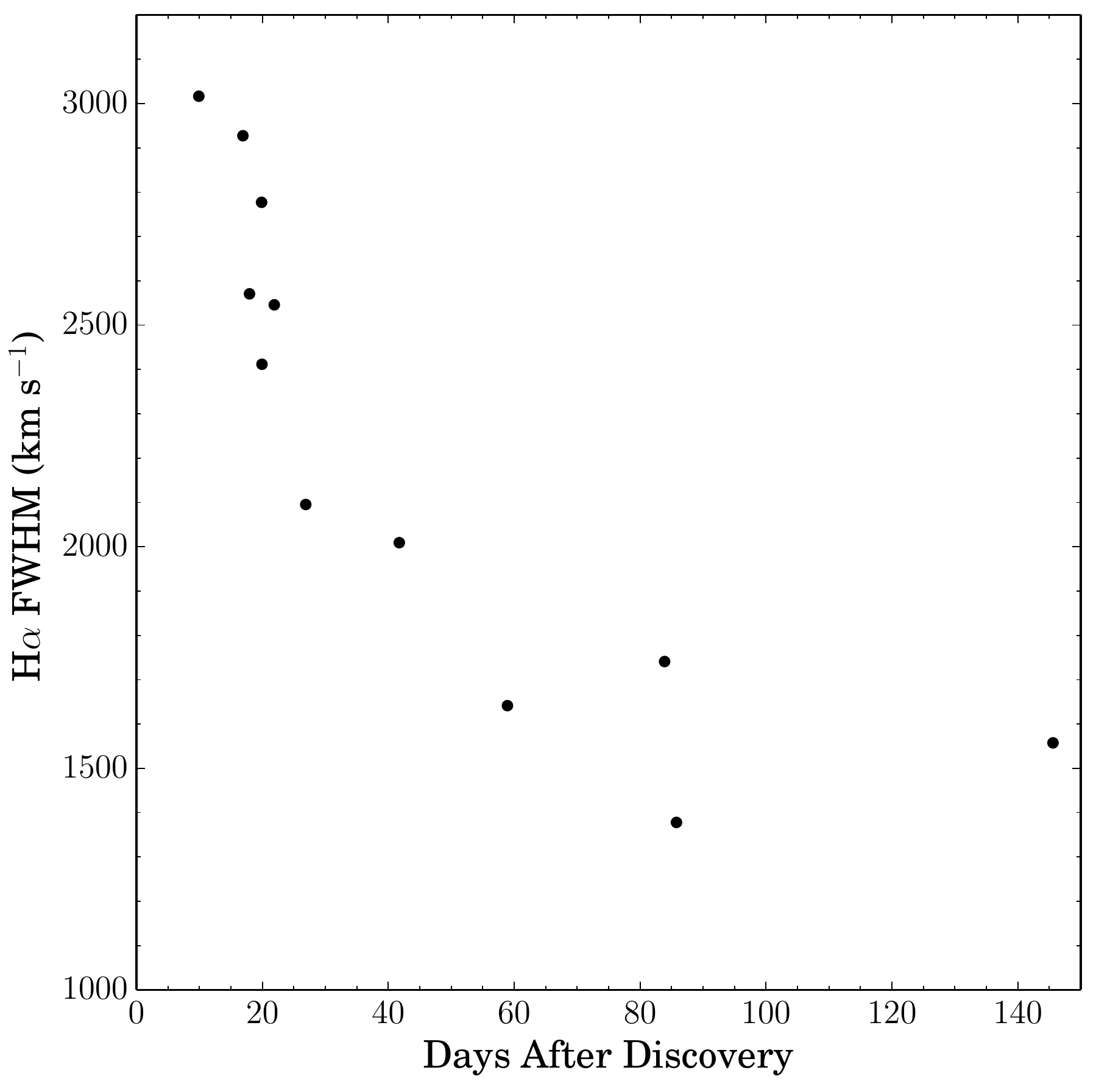}}}
\caption{Evolution of the H$\alpha$ line width. As the luminosity of the line also decreases with time (see Figure~\ref{fig:line_lum}), the line is becoming narrower as the luminosity decreases. This is the opposite of what is seen in reverberation mapping studies of quasars.}
\label{fig:line_width}
\end{figure}

Finally, we examine the evolution of the emission line widths, shown in Figure~\ref{fig:line_width} for H$\alpha$. As the luminosity of the transient is decreasing, so too is the line width. This is the opposite of what is seen in reverberation mapping studies of quasars, where estimates of the black hole mass $M_{BH} \propto \Delta v^2 L^{1/2}$ remain roughly constant because the luminosity decreases as the line width broadens \citep[e.g.,][]{peterson04,denney09}. Physically, this is believed to result from the fact that if the luminosity drops, gas at larger distances and lower average velocities recombines, leading to an increase in the line width.

Simple estimates based on the line luminosities and light travel times imply that the densities of the ionized regions of {\name} are also high enough to make the recombination times negligible. This means it is unlikely that the narrowing of the lines is due to a finely-tuned outward density gradient allowing the higher velocity material at smaller radii to recombine faster while also making the line width shrink with time. The fast recombination times also make it difficult to explain the decreasing line widths as simply being due to a continuing expansion of the Stromgren sphere even as the luminosity is decreasing.

No TDE has been caught early enough to make these measurements, but there should be temporal lags between the rise of the UV emission and the formation of the broad lines, as is seen in reverberation mapping of AGN \citep[e.g.,][]{peterson04}. At late times, these effects are still present, but seem an unlikely explanation for the observed line width evolution, as the temporal smoothing of the changes becomes larger when the delays are long.

A final possibility is that the changes represent evolution in the density distribution of the ionized gas. While there is no reason to expect rapid, large scale gas redistributions in a normal AGN, such changes seem plausible during a TDE. This is presumably not a large scale redistribution, since the time scale for a significant change in radius is $ t \sim 50 M_{BH7}v_3^{-3}$~years, where $M_{BH7}=(10^7{\msun})M_{BH}$ and $v_3=(3,000~\textrm{km~s}^{-1})v$. It would not be surprising, however, to have significant evolution in the mean density at a given radius given the nature of a TDE. For example, if the line emission is dominated by a dense gas phase, but the average density of this phase is decreasing, the total line emission diminishes in proportion to the density. Given the quality of the spectra of {\name}, we do not attempt any quantitative analysis, but these questions suggest a need for higher quality spectra in future studies of TDEs.

The spectra of {\name} seem to be consistent with both those of ASASSN-14ae and those of other TDE candidates in literature, showing strong emission at bluer wavelengths that fades steadily over time and strong Balmer and helium emission features in all epochs. These similarities, as well as the fact that these spectra do not seem to resemble those of type II supernovae or AGN, provide strong evidence for a TDE interpretation for {\name}.


\section{TDE Rates}
\label{sec:rates}

Due to the small number of candidate TDEs, the rate of stellar tidal disruptions by SMBHs is not particularly well-understood. This rate may be important, however, as it can be used to study the orbits of stars around SMBHs and the origin of relativistic TDEs that have been observed with {\swift} \citep{velzen14}. Furthermore, if the rate is high enough, it may even play an important role in the growth of SMBHs \citep[e.g.,][]{magorrian99}. Van Velzen \& Farrar (2014) used four TDE detections (two from SDSS and two from Pan-STARRS) and a search of SDSS Stripe 82 galaxies to estimate a TDE rate of $(1.5 - 2.0)_{-1.3}^{+2.7} \times 10^{-5}~{\rm yr}^{-1}$ per galaxy. In \citet{holoien14b} we used estimates for the local density of black holes from \citet{shankar13} and an observable volume of $3\times10^7$~Mpc$^3$ for ASAS-SN to estimate that ASAS-SN would discover $0.3 - 3$ TDEs per year, assuming a 50\% detection efficiency. If we take our estimate of 50\% detection efficiency to be true, the fact that ASAS-SN discovered 2 TDE candidates in 2014 would imply that the actual rate of tidal disruptions is actually much higher than the estimate from \citet{velzen14}, possibly as high as $\sim10^{-4}$~yr$^{-1}$ per galaxy. However, the assumption of a 50\% detection efficiency is arbitrary, and as the rate estimate depends strongly on this quantity, a more realistic simulation of the ASAS-SN detection efficiency is needed.

To make a more quantitative estimate of the TDE rate, we used SDSS galaxies, which includes the hosts of both of our TDE candidates. We selected all galaxies from SDSS DR9 with $0.01 < z < 0.10$, no duplicate spectra, and with photometric data. We randomly assigned all $N_g=348853$ galaxies an integer code from $1$ to $100$ and extracted aperture light curves for the centers of a randomly selected 4\% of these galaxies. We analyzed all light curves from 2014 January 1 to 2015 May 1.

We modeled the peak of the light curve, which will dominate any magnitude limited detections, as $M_V = V_{peak} + (t-t_0)^2/t_1^2$ where $V_{peak}$ is the peak absolute magnitude, $t_0$ is the time of peak, and $t_1$ is the time to decay by one magnitude from the peak.  For each galaxy, these were converted to  apparent magnitudes using a quadratic fit to the luminosity distance for an $H_0=70$~km~s$^{-1}$~Mpc$^{-1}$, $\Omega_0=0.3$, $\Omega_\Lambda=0.7$ flat cosmological model combined with the Galactic extinction in each sightline.  For each galaxy, $N_t=1000$ trial values of $t_0$ were drawn from one month before the start of the light curve to one month afterward for a total time span of $\Delta t_m$, and the trial source was viewed as detected if there would be two signal-to-noise ratio $S/N>7.5$ detections within one week given the model and the estimated noise in the actual light curve for the center of the galaxy. This criterion detects both of our observed candidates. With these definitions, the survey time over which we would detect a TDE in any single galaxy is $\Delta t=\Delta t_m N_d/N_t$, where $N_d$ is the number of trial detections. Averaging this over all the trial galaxies gives us a mean survey time per galaxy $\langle \Delta t \rangle$, leading to a rate estimate of
\begin{equation}
           r = { N_{TDE}  \over N_g \langle \Delta t \rangle }.
\end{equation}
The results as a function of $V_{peak}$ and $t_1$ are presented in Table~\ref{tab:survey}.  We have not included results for transients fainter than $V_{peak}=-18$~mag because they begin to be significantly affected by the $z>0.01$ redshift cutoff.  

Using ASASSN-14ae \citep{holoien14b} and PS1-10jh \citep{gezari12b} as examples of ``typical'' TDEs, we assume that typical TDEs have $V_{peak} \simeq -19$ to $-20$ and decay by one magnitude in $25$ to $40$~days.  In Table~\ref{tab:survey}, we see that the time scale $t_1$ has only a modest affect on the effective survey time once $t_1>10$~days, while the peak magnitude has an enormous effect because it controls the effective survey volume. If we assume that TDEs are uniformly distributed over the $V_{peak}=-19$ and $-20$~mag bins for $t_1=20$, $30$ and $40$~days, we find that $N_g \langle \Delta t \rangle \simeq 49,000$~years implying an average TDE rate of \edit{$r \simeq 4.1 \times 10^{-5}~{\rm yr}^{-1}$} per galaxy given $N_{TDE}=2$.  The Poisson uncertainties correspond to a 90\% confidence range of $(2.2 - 17.0) \times 10^{-5}~{\rm yr}^{-1}$ per galaxy, and these probably dominate over the systematic uncertainties. For example, the fractional shifts from taking any of the cases averaged over to yield the estimate of $N_g \langle \Delta t\rangle$ are significantly smaller than the Poisson uncertainties. This rate estimate is higher than the rate of $(1.5 - 2.0)_{-1.3}^{+2.7} \times 10^{-5}~{\rm yr}^{-1}$ per galaxy found by \citet{velzen14}, though the two estimates are consistent given the uncertainties. If we lower the threshold to two observations with $S/N>5$, the rate estimate drops by a factor of 1.5 and is closer to the estimate of \citet{velzen14}, but the predicted magnitude distribution of Type Ia SN for this threshold is somewhat fainter than is observed. If we raise the threshold to $S/N>10$, the rate estimate rises by a factor of 1.4 and is closer to theoretical predictions \citep[e.g.,][]{stone14,metzger15}, but the predicted magnitude distribution of Type Ia SN is somewhat brighter than is observed and the two TDEs no longer satisfy the detection criterion.


\begin{table}
\centering
\caption{Detection Statistics}
\renewcommand{\arraystretch}{1.2}
\begin{tabular}{ccrr}
\hline
$V_{peak}$ & $t_1$ (days) & $\langle\Delta t\rangle$ (days) & $N_g \Delta t$ (years) \\
\hline
   $-18$    &10 &  4.88 & \edit{4661} \\
          &20 & 6.31 & \edit{6028} \\
          &30 & 7.11 & \edit{6793} \\
          &40 & 7.70 & \edit{7351} \\
   $-19$    &10 & 20.51 & \edit{19588} \\
          &20 & 25.48 & \edit{24331} \\
          &30 & 28.53 & \edit{27251} \\
          &40 & 30.68 & \edit{29307} \\
   $-20$    &10 & 55.78 & \edit{53280} \\
          &20 & 67.77 & \edit{64730} \\
          &30 & 75.13 & \edit{71753} \\
          &40 & 80.60 & \edit{76980} \\
   $-21$    &10 & 131.82 & \edit{125903} \\
          &20 & 159.56 & \edit{152400} \\
          &30 & 176.58 & \edit{168655} \\
          &40 & 189.30 & \edit{180803} \\
\hline
\end{tabular}

\medskip
\raggedright
\noindent  For a given transient peak $V_{peak}$ and time to decay one magnitude $t_1$, $\Delta t$ is the average number of days per galaxy in which the transient would be detected, leading to a total survey time of $N_g \Delta t$ where $N_g=348853$ is the total number of $0.01 < z < 0.1$ SDSS galaxies we considered.
\label{tab:survey}
\end{table}


\section{Discussion}
\label{sec:disc}

{\name}, discovered by ASAS-SN on 2014 November 11, had a position consistent to within $0.17\pm0.21$~arcseconds of the center of {\galname} and a peak absolute $V$-band magnitude of $M_V\sim-19$. Follow-up observations indicate that it is not consistent with either a supernova or a normal AGN outburst. Conversely, it shows many similarities with other TDE candidates discovered by optical surveys. {\name} has remained bright in the UV and blue optical filters even six months after detection, and the best-fit blackbody temperature has remained roughly constant at $T\sim35,000$~K for the duration of the outburst while the luminosity has declined at a steady rate best fit by an exponential decay curve. Spectra of {\name} show a strong blue continuum and broad Balmer and helium emission features in all epochs, and do not show the spectral evolution expected of supernovae or AGN. Such features are characteristic of TDE candidates such as those discovered by ASAS-SN \citep[ASASSN-14ae\edit{,}][]{holoien14b} and iPTF \citep{arcavi14}, which leads us to the conclusion that {\name} was likely a TDE as well. Its proximity ($z=0.0206$) makes it the closest TDE candidate discovered at optical wavelengths to date, and it is also the first TDE candidate discovered at optical wavelengths to exhibit both UV/optical emission as well as associated X-ray emission.

Archival spectroscopy and photometry, as well as SED fitting, indicate that the transient host galaxy {\galname} has undergone a change in its star formation history within the last $\sim1$ Gyr and has a stellar population dominated by A stars, implying that it is a post-starburst galaxy. Previous work by \citet{arcavi14} indicated that TDE candidates may prefer such hosts, and this possibility will be further explored in future work on ASAS-SN TDE host galaxies (Dong et al., \emph{in prep.}). The host does not show signs of recent star formation, and while the detection of [\ion{O}{3}] emission and radio emission may imply that it hosts a weak AGN, its mid-IR colors from WISE are inconsistent with significant AGN activity. 

{\name} has significantly more HeII emission than ASASSN-14ae, so much so that the HeII Stromgren sphere is likely of comparable size to the H Stromgren sphere. This is consistent with the diversity of He and H line strengths noted by \citet{arcavi14}.  The observed soft UV SEDs, line luminosities, and X-ray properties of the two ASAS-SN TDE candidates also provide a natural explanation of the diversity. {\name} has both a harder soft UV continuum and significant X-ray flux, which implies a stronger hard/ionizing UV continuum. This is then observed indirectly through the greater ratio of He to H emission in the spectra. While the unknown covering fractions mean that the observed line luminosities only set lower bounds on the hard UV continuum, their ratios likely provide reasonable estimates of the spectral slopes because it is difficult to make the H and He line emission regions enormously different.

The evolution of the line widths is also interesting, as discussed in more detail in \S~\ref{sec:specanal}. For the ``quiescent'' environments of reverberation mapped AGN, decreases in luminosity are accompanied by increases in line widths \citep[e.g.][]{peterson04,denney09}. For {\name} we see the opposite: as the transient fades, the line widths become narrower. One likely explanation for this behavior is that the line width evolution is due to more rapid evolution of the density distribution closer to the black hole. We do not mean in the sense of large scale changes in the amount of mass at a given radius -- for the typical velocities of the lines (1,000-3,000~km~s$^{-1}$), this occurs relatively slowly (years to decades) if the velocities are virialized. However, rearrangements of the gas that change the mean emission measure can occur much more rapidly, so the line evolution may provide a probe of the evolution of the ``clumpiness'' of the material on these velocity scales.

Both of these issues are worth exploring in more detail, but they also need higher quality spectra than we have available for {\name}. Spectra with significantly higher signal-to-noise ratios obtained at a higher cadence are needed to closely track the evolution of the structure of the emission lines. In such a close study it will be important to bear in mind the effects of light travel times (days to months), potentially with the possibility of essentially carrying out a reverberation mapping study of a TDE.  This would also require a high cadence continuum light curve including the onset of the transient.

We used SDSS galaxies including the hosts of the two TDE candidates discovered by ASAS-SN to estimate the rate of tidal disruptions in galaxies with redshift $0.01 < z < 0.10$. We extracted aperture light curves from the centers of randomly selected galaxies to model the completeness. Assuming that TDEs are uniformly distributed between $V_{peak}=-19$ and $V_{peak}=-20$ and have a characteristic decay time between $20$, $30$ and $40$~days (roughly consistent with most TDE candidates in literature), we find a 90\% confidence range of $r=(2.2 - 17.0) \times 10^{-5}~{\rm yr}^{-1}$ per galaxy given the two ASAS-SN TDE candidates discovered between 2014 January 1 and 2015 May 1. This rate is in agreement with the rate found by \citet{velzen14} in their analysis of SDSS TDE candidates, though our average rate estimate is higher, and will be further refined in future work by Dong et al. (\emph{in prep.}).

We also note that if we carry out a similar rate analysis for Type Ia SNe in ASAS-SN, we find a rate consistent with that found by the LOSS survey \citep{li11}, suggesting that the overall analysis is reasonably robust. A detailed analysis of Type Ia rates in ASAS-SN will be carried out in a future publication by Holoien et al. (\emph{in prep.}).

TDEs and Type Ia SNe have broadly similar peak magnitudes and time scales at peak, so we would naively expect the two source types to have similar selection effects and hence ratios of event numbers between surveys. For ASAS-SN, this ratio is approximately 1 TDE for every 70 Type Ia SNe (2 and 135, respectively). The ratios for PTF (\citealt{arcavi14,rau09}), Pan-STARRS and SDSS (\citealt{gezari12b,chornock14,velzen14}) are are 1 TDE for  every 550, 1000, and 1050 SNe Ia, respectively.\footnote{This is based on 3, 2 and 2 TDE candidates in samples of 550, $>2000$ (\citealt{jones15}), and $\sim 2100$ (\citealt{sako14}) Type Ia SNe.} Unless we have been unusually lucky (for the PTF/Pan-STARRS/SDSS ratios, we would have a 1-2\% probability of finding two TDEs), this comparison suggests the (testable) possibility that the completeness of TDE searches in ASAS-SN is markedly higher than in prior surveys. This offers the exciting possibility that a much larger population of TDEs can be identified in existing surveys, although the sample found by ASAS-SN will remain the most useful for detailed study due to their intrinsic brightness. It remains to be seen, however, if this will help explain the gap between observed and theoretical TDE rate estimates (see \citealt{metzger15}).

{\name} is the closest TDE candidate ever discovered at optical wavelengths, and it has been extensively observed for over 6 months since discovery, resulting in an unprecedented data set spanning optical to X-ray wavelengths. Given the design of the ASAS-SN survey, future TDE candidates that we discover will be similarly easy to observe with a variety of telescopes and instruments, allowing us to develop a catalog of well-studied TDE candidates that can be used for population studies and to study the early and late-time behaviors of these transients, which cannot be done with TDEs discovered at higher redshifts. With a planned expansion to 8 cameras in mid-2015, ASAS-SN will be an even more powerful tool for the discovery and study of TDEs and other bright transients in the future.

\section*{Acknowledgments}

The authors thank PI Neil Gehrels and the {\swift} ToO team for promptly approving and executing our observations. We thank D. Mudd for assistance with data analysis and R. Pogge for discussion. We thank LCOGT and its staff for their continued support of ASAS-SN.

Development of ASAS-SN has been supported by NSF grant AST-0908816 and the Center for Cosmology and AstroParticle Physics at the Ohio State University. ASAS-SN is supported in part by Mt. Cuba Astronomical Foundation.

TW-SH is supported by the DOE Computational Science Graduate Fellowship, grant number DE-FG02-97ER25308. Support for JLP is in part provided by FONDECYT through the grant 1151445 and by the Ministry of Economy, Development, and Tourism's Millennium Science Initiative through grant IC120009, awarded to The Millennium Institute of Astrophysics, MAS. SD is supported by the Strategic Priority Research Program-``The Emergence of Cosmological Structures of the Chinese Academy of Sciences (Grant No. XDB09000000)''. BJS is supported by NASA through Hubble Fellowship grant HST-HF-51348.001 awarded by the Space Telescope Science Institute, which is operated by the Association of Universities for Research in Astronomy, Inc., for NASA, under contract NAS 5-26555. JFB is supported by NSF grant PHY-1404311. PRW is supported by the Laboratory Directed Research and Development program at LANL.

This research has made use of the XRT Data Analysis Software (XRTDAS) developed under the responsibility of the ASI Science Data Center (ASDC), Italy. At Penn State the NASA {\swift} program is support through contract NAS5-00136.

This research uses data obtained through the Telescope Access Program(TAP), which has also been funded by the aforementioned Strategic Priority Research Program and the Special Fund for Astronomy from the Ministry of Finance.

Observations made with the NASA Galaxy Evolution Explorer (GALEX) were used in the analyses presented in this manuscript. Some of the data presented in this paper were obtained from the Mikulski Archive for Space Telescopes (MAST). STScI is operated by the Association of Universities for Research in Astronomy, Inc., under NASA contract NAS5-26555. Support for MAST for non-HST data is provided by the NASA Office of Space Science via grant NNX13AC07G and by other grants and contracts.

The Liverpool Telescope is operated on the island of La Palma by Liverpool John Moores University in the Spanish Observatorio del Roque de los Muchachos of the Instituto de Astrofisica de Canarias with financial support from the UK Science and Technology Facilities Council.

This research was made possible through the use of the AAVSO Photometric All-Sky Survey (APASS), funded by the Robert Martin Ayers Sciences Fund.

This research has made use of data provided by Astrometry.net \citep{barron08}.

Funding for SDSS-III has been provided by the Alfred P. Sloan Foundation, the Participating Institutions, the National Science Foundation, and the U.S. Department of Energy Office of Science. The SDSS-III web site is http://www.sdss3.org/.

SDSS-III is managed by the Astrophysical Research Consortium for the Participating Institutions of the SDSS-III Collaboration including the University of Arizona, the Brazilian Participation Group, Brookhaven National Laboratory, Carnegie Mellon University, University of Florida, the French Participation Group, the German Participation Group, Harvard University, the Instituto de Astrofisica de Canarias, the Michigan State/Notre Dame/JINA Participation Group, Johns Hopkins University, Lawrence Berkeley National Laboratory, Max Planck Institute for Astrophysics, Max Planck Institute for Extraterrestrial Physics, New Mexico State University, New York University, Ohio State University, Pennsylvania State University, University of Portsmouth, Princeton University, the Spanish Participation Group, University of Tokyo, University of Utah, Vanderbilt University, University of Virginia, University of Washington, and Yale University.

This publication makes use of data products from the Two Micron All Sky Survey, which is a joint project of the University of Massachusetts and the Infrared Processing and Analysis Center/California Institute of Technology, funded by the National Aeronautics and Space Administration and the National Science Foundation.

This publication makes use of data products from the Wide-field Infrared Survey Explorer, which is a joint project of the University of California, Los Angeles, and the Jet Propulsion Laboratory/California Institute of Technology, funded by the National Aeronautics and Space Administration.

This research has made use of the NASA/IPAC Extragalactic Database (NED), which is operated by the Jet Propulsion Laboratory, California Institute of Technology, under contract with the National Aeronautics and Space Administration.

\bibliographystyle{mn2e}
\bibliography{bibliography}


\appendix
\section{Follow-up Photometry and Line Luminosities}
All follow-up photometry and line luminosities are presented in Table~\ref{tab:phot} and Table~\ref{tab:line_lum} below, respectively. Photometry is presented in the natural system for each filter: $ugriz$ magnitudes are in the AB system, while {\swift} filter magnitudes are in the Vega system.


\begin{table*}
\begin{minipage}{\textwidth}
\centering
\caption{Photometric data of {\name}.\hfill}
\renewcommand{\arraystretch}{1.2}
\begin{tabular}{cccc|cccc}
\hline
MJD & Magnitude &  Filter & Telescope & MJD & Magnitude &  Filter & Telescope\\
\hline
57007.258 & 15.013 0.013 & $z$ & LT & 57017.529 & 15.284 0.012 & $i$ & LCOGT\\ 
57008.280 & 15.013 0.014 & $z$ & LT & 57017.532 & 15.296 0.012 & $i$ & LCOGT\\ 
57009.275 & 15.102 0.013 & $z$ & LT & 57018.163 & 15.265 0.010 & $i$ & LT\\ 
57011.227 & 15.087 0.014 & $z$ & LT & 57019.155 & 15.270 0.009 & $i$ & LT\\ 
57012.211 & 15.028 0.014 & $z$ & LT & 57020.319 & 15.284 0.013 & $i$ & LCOGT\\ 
57014.176 & 15.002 0.018 & $z$ & LT & 57020.322 & 15.273 0.012 & $i$ & LCOGT\\ 
57016.213 & 15.027 0.013 & $z$ & LT & 57021.155 & 15.274 0.009 & $i$ & LT\\ 
57017.205 & 15.120 0.014 & $z$ & LT & 57021.701 & 15.316 0.012 & $i$ & LCOGT\\ 
57018.164 & 15.110 0.016 & $z$ & LT & 57021.703 & 15.287 0.012 & $i$ & LCOGT\\ 
57019.156 & 15.095 0.015 & $z$ & LT & 57024.214 & 15.302 0.010 & $i$ & LT\\ 
57020.153 & 15.038 0.022 & $z$ & LT & 57025.277 & 15.286 0.008 & $i$ & LT\\ 
57021.156 & 15.076 0.014 & $z$ & LT & 57025.692 & 15.287 0.018 & $i$ & LCOGT\\ 
57023.131 & 15.035 0.014 & $z$ & LT & 57026.122 & 15.252 0.017 & $i$ & LT\\ 
57024.214 & 15.119 0.015 & $z$ & LT & 57028.042 & 15.310 0.019 & $i$ & LCOGT\\ 
57025.277 & 15.114 0.011 & $z$ & LT & 57029.145 & 15.286 0.013 & $i$ & LT\\ 
57026.122 & 15.033 0.018 & $z$ & LT & 57029.679 & 15.296 0.002 & $i$ & LCOGT\\ 
57027.120 & 15.042 0.017 & $z$ & LT & 57029.682 & 15.299 0.020 & $i$ & LCOGT\\ 
57028.200 & 15.110 0.013 & $z$ & LT & 57032.031 & 15.288 0.024 & $i$ & LCOGT\\ 
57029.119 & 15.043 0.016 & $z$ & LT & 57037.237 & 15.321 0.005 & $i$ & LT\\ 
57029.145 & 15.023 0.016 & $z$ & LT & 57038.211 & 15.307 0.005 & $i$ & LT\\ 
57037.238 & 15.113 0.008 & $z$ & LT & 57041.007 & 15.316 0.016 & $i$ & LCOGT\\ 
57038.212 & 15.115 0.007 & $z$ & LT & 57041.010 & 15.319 0.016 & $i$ & LCOGT\\ 
57039.157 & 15.043 0.007 & $z$ & LT & 57043.003 & 15.343 0.016 & $i$ & LCOGT\\ 
57041.138 & 15.056 0.008 & $z$ & LT & 57043.006 & 15.329 0.016 & $i$ & LCOGT\\ 
57042.143 & 15.059 0.008 & $z$ & LT & 57044.152 & 15.314 0.005 & $i$ & LT\\ 
57044.153 & 15.131 0.006 & $z$ & LT & 57045.111 & 15.298 0.012 & $i$ & LCOGT\\ 
57046.083 & 15.149 0.007 & $z$ & LT & 57045.113 & 15.321 0.012 & $i$ & LCOGT\\ 
57048.079 & 15.159 0.007 & $z$ & LT & 57046.082 & 15.318 0.005 & $i$ & LT\\ 
57051.083 & 15.161 0.017 & $z$ & LT & 57047.048 & 15.342 0.013 & $i$ & LCOGT\\ 
57053.056 & 15.142 0.008 & $z$ & LT & 57047.050 & 15.303 0.013 & $i$ & LCOGT\\ 
57055.048 & 15.129 0.009 & $z$ & LT & 57048.078 & 15.334 0.005 & $i$ & LT\\ 
57057.082 & 15.120 0.009 & $z$ & LT & 57050.994 & 15.320 0.016 & $i$ & LCOGT\\ 
57059.043 & 15.157 0.010 & $z$ & LT & 57051.082 & 15.356 0.016 & $i$ & LT\\ 
57065.160 & 15.135 0.007 & $z$ & LT & 57053.054 & 15.321 0.006 & $i$ & LT\\ 
57068.143 & 15.136 0.007 & $z$ & LT & 57055.047 & 15.311 0.007 & $i$ & LT\\ 
56989.467 & 15.291 0.053 & $i$ & LCOGT & 57057.081 & 15.299 0.008 & $i$ & LT\\ 
57003.460 & 15.213 0.016 & $i$ & LCOGT & 57059.042 & 15.276 0.009 & $i$ & LT\\ 
57003.462 & 15.178 0.015 & $i$ & LCOGT & 57065.159 & 15.306 0.005 & $i$ & LT\\ 
57004.468 & 15.223 0.019 & $i$ & LCOGT & 57068.142 & 15.316 0.005 & $i$ & LT\\ 
57004.471 & 15.253 0.019 & $i$ & LCOGT & 56989.465 & 15.404 0.036 & $r$ & LT\\ 
57006.414 & 15.214 0.015 & $i$ & LCOGT & 57007.257 & 15.375 0.008 & $r$ & LT\\ 
57006.417 & 15.203 0.016 & $i$ & LCOGT & 57008.279 & 15.447 0.008 & $r$ & LT\\ 
57006.438 & 15.260 0.038 & $i$ & LCOGT & 57009.274 & 15.453 0.007 & $r$ & LT\\ 
57006.441 & 15.234 0.033 & $i$ & LCOGT & 57011.226 & 15.459 0.007 & $r$ & LT\\ 
57006.471 & 15.260 0.013 & $i$ & LCOGT & 57012.210 & 15.396 0.007 & $r$ & LT\\ 
57006.473 & 15.279 0.013 & $i$ & LCOGT & 57014.174 & 15.402 0.009 & $r$ & LT\\ 
57008.280 & 15.265 0.010 & $i$ & LT & 57016.212 & 15.415 0.007 & $r$ & LT\\ 
57009.274 & 15.259 0.009 & $i$ & LT & 57017.204 & 15.484 0.008 & $r$ & LT\\ 
57011.227 & 15.250 0.009 & $i$ & LT & 57018.163 & 15.480 0.008 & $r$ & LT\\ 
57014.082 & 15.255 0.014 & $i$ & LCOGT & 57019.155 & 15.461 0.008 & $r$ & LT\\ 
57014.089 & 15.233 0.015 & $i$ & LCOGT & 57020.152 & 15.451 0.012 & $r$ & LT\\ 
57014.093 & 15.252 0.015 & $i$ & LCOGT & 57021.154 & 15.485 0.008 & $r$ & LT\\ 
57016.076 & 15.298 0.012 & $i$ & LCOGT & 57024.213 & 15.506 0.009 & $r$ & LT\\ 
57016.079 & 15.271 0.020 & $i$ & LCOGT & 57025.276 & 15.503 0.008 & $r$ & LT\\ 
57017.205 & 15.277 0.009 & $i$ & LT & 57026.121 & 15.503 0.019 & $r$ & LT\\ 
\hline
\end{tabular}
\label{tab:phot}
\end{minipage}
\end{table*}

\begin{table*}
\begin{minipage}{\textwidth}
\centering
\renewcommand{\arraystretch}{1.2}
\begin{tabular}{cccc|cccc}
\hline
MJD & Magnitude &  Filter & Telescope & MJD & Magnitude &  Filter & Telescope\\
\hline
57028.199 & 15.511 0.012 & $r$ & LT & 57008.279 & 15.830 0.009 & $g$ & LT\\ 
57029.144 & 15.462 0.012 & $r$ & LT & 57009.273 & 15.845 0.008 & $g$ & LT\\ 
57037.236 & 15.518 0.004 & $r$ & LT & 57011.226 & 15.840 0.008 & $g$ & LT\\ 
57038.210 & 15.526 0.004 & $r$ & LT & 57012.209 & 15.856 0.009 & $g$ & LT\\ 
57039.155 & 15.472 0.004 & $r$ & LT & 57014.174 & 15.856 0.011 & $g$ & LT\\ 
57041.136 & 15.497 0.005 & $r$ & LT & 57016.073 & 15.887 0.010 & $g$ & LCOGT\\ 
57042.141 & 15.493 0.004 & $r$ & LT & 57016.211 & 15.857 0.008 & $g$ & LT\\ 
57044.151 & 15.535 0.004 & $r$ & LT & 57017.203 & 15.887 0.008 & $g$ & LT\\ 
57046.081 & 15.539 0.004 & $r$ & LT & 57017.526 & 15.910 0.008 & $g$ & LCOGT\\ 
57048.077 & 15.546 0.004 & $r$ & LT & 57018.162 & 15.867 0.009 & $g$ & LT\\ 
57051.081 & 15.518 0.015 & $r$ & LT & 57019.154 & 15.873 0.008 & $g$ & LT\\ 
57053.053 & 15.537 0.007 & $r$ & LT & 57020.151 & 15.916 0.015 & $g$ & LT\\ 
57055.046 & 15.551 0.008 & $r$ & LT & 57020.317 & 15.905 0.010 & $g$ & LCOGT\\ 
57057.080 & 15.558 0.009 & $r$ & LT & 57021.154 & 15.900 0.008 & $g$ & LT\\ 
57059.041 & 15.550 0.011 & $r$ & LT & 57021.698 & 15.931 0.011 & $g$ & LCOGT\\ 
57065.158 & 15.540 0.004 & $r$ & LT & 57023.130 & 15.946 0.017 & $g$ & LT\\ 
57068.141 & 15.548 0.004 & $r$ & LT & 57024.213 & 15.970 0.012 & $g$ & LT\\ 
56991.431 & 15.570 0.061 & $V$ & {\swift} & 57025.275 & 15.930 0.008 & $g$ & LT\\ 
56993.888 & 15.670 0.061 & $V$ & {\swift} & 57025.687 & 15.900 0.034 & $g$ & LCOGT\\ 
56995.292 & 15.570 0.061 & $V$ & {\swift} & 57026.120 & 15.889 0.032 & $g$ & LT\\ 
56998.274 & 15.620 0.061 & $V$ & {\swift} & 57027.118 & 15.942 0.024 & $g$ & LT\\ 
57001.608 & 15.470 0.051 & $V$ & {\swift} & 57028.040 & 15.956 0.026 & $g$ & LCOGT\\ 
57004.264 & 15.700 0.100 & $V$ & {\swift} & 57028.198 & 15.960 0.016 & $g$ & LT\\ 
57007.266 & 15.650 0.061 & $V$ & {\swift} & 57029.117 & 15.948 0.021 & $g$ & LT\\ 
57010.803 & 15.630 0.061 & $V$ & {\swift} & 57029.144 & 15.956 0.018 & $g$ & LT\\ 
57013.067 & 15.710 0.061 & $V$ & {\swift} & 57029.677 & 15.958 0.030 & $g$ & LCOGT\\ 
57016.062 & 15.690 0.061 & $V$ & {\swift} & 57037.235 & 15.963 0.004 & $g$ & LT\\ 
57019.515 & 15.720 0.071 & $V$ & {\swift} & 57038.209 & 15.966 0.004 & $g$ & LT\\ 
57022.718 & 15.690 0.061 & $V$ & {\swift} & 57039.154 & 15.968 0.004 & $g$ & LT\\ 
57029.319 & 15.690 0.081 & $V$ & {\swift} & 57041.004 & 15.996 0.014 & $g$ & LCOGT\\ 
57033.113 & 15.780 0.081 & $V$ & {\swift} & 57041.135 & 15.985 0.004 & $g$ & LT\\ 
57036.108 & 15.890 0.081 & $V$ & {\swift} & 57042.140 & 15.975 0.004 & $g$ & LT\\ 
57039.036 & 15.870 0.071 & $V$ & {\swift} & 57043.000 & 15.977 0.014 & $g$ & LCOGT\\ 
57042.230 & 15.940 0.091 & $V$ & {\swift} & 57044.150 & 15.990 0.004 & $g$ & LT\\ 
57045.560 & 15.790 0.071 & $V$ & {\swift} & 57045.108 & 15.986 0.010 & $g$ & LCOGT\\ 
57048.756 & 15.770 0.081 & $V$ & {\swift} & 57046.080 & 15.993 0.004 & $g$ & LT\\ 
57051.468 & 15.750 0.081 & $V$ & {\swift} & 57047.042 & 15.985 0.016 & $g$ & LCOGT\\ 
57054.007 & 15.750 0.081 & $V$ & {\swift} & 57048.076 & 16.005 0.004 & $g$ & LT\\ 
57057.530 & 15.770 0.061 & $V$ & {\swift} & 57051.080 & 15.974 0.020 & $g$ & LT\\ 
57059.798 & 15.830 0.091 & $V$ & {\swift} & 57053.052 & 15.967 0.011 & $g$ & LT\\ 
57065.852 & 15.740 0.071 & $V$ & {\swift} & 57055.045 & 16.036 0.013 & $g$ & LT\\ 
57068.781 & 15.810 0.061 & $V$ & {\swift} & 57057.079 & 16.001 0.014 & $g$ & LT\\ 
57071.710 & 15.750 0.061 & $V$ & {\swift} & 57059.040 & 16.073 0.017 & $g$ & LT\\ 
57074.848 & 15.800 0.081 & $V$ & {\swift} & 57065.157 & 15.985 0.005 & $g$ & LT\\ 
57077.576 & 15.850 0.071 & $V$ & {\swift} & 57068.130 & 16.006 0.003 & $g$ & LT\\ 
57080.891 & 15.700 0.110 & $V$ & {\swift} & 57068.140 & 16.006 0.004 & $g$ & LT\\ 
57086.887 & 15.850 0.071 & $V$ & {\swift} & 57070.125 & 16.021 0.003 & $g$ & LT\\ 
57089.350 & 15.760 0.061 & $V$ & {\swift} & 57077.222 & 16.069 0.005 & $g$ & LT\\ 
57132.462 & 15.710 0.081 & $V$ & {\swift} & 57079.071 & 16.062 0.003 & $g$ & LT\\ 
56989.463 & 15.752 0.041 & $g$ & LCOGT & 57085.032 & 16.055 0.008 & $g$ & LT\\ 
57003.457 & 15.715 0.014 & $g$ & LCOGT & 57093.076 & 16.082 0.004 & $g$ & LT\\ 
57004.465 & 15.740 0.016 & $g$ & LCOGT & 57095.071 & 16.081 0.003 & $g$ & LT\\ 
57006.412 & 15.782 0.016 & $g$ & LCOGT & 57097.994 & 16.095 0.003 & $g$ & LT\\ 
57006.436 & 15.797 0.056 & $g$ & LCOGT & 57109.927 & 16.116 0.005 & $g$ & LT\\ 
57006.468 & 15.794 0.011 & $g$ & LCOGT & 57112.017 & 16.112 0.009 & $g$ & LT\\ 
57007.256 & 15.798 0.010 & $g$ & LT & 57113.968 & 16.102 0.011 & $g$ & LT\\ 
\hline
\end{tabular}
\end{minipage}
\end{table*}

\begin{table*}
\begin{minipage}{\textwidth}
\centering
\renewcommand{\arraystretch}{1.2}
\begin{tabular}{cccc|cccc}
\hline
MJD & Magnitude &  Filter & Telescope & MJD & Magnitude &  Filter & Telescope\\
\hline
57117.977 & 16.110 0.009 & $g$ & LT & 57020.153 & 16.716 0.134 & $u$ & LT\\ 
57121.021 & 16.099 0.004 & $g$ & LT & 57021.156 & 16.744 0.090 & $u$ & LT\\ 
57122.959 & 16.115 0.003 & $g$ & LT & 57024.215 & 16.842 0.069 & $u$ & LT\\ 
57124.975 & 16.114 0.003 & $g$ & LT & 57025.278 & 16.842 0.052 & $u$ & LT\\ 
57127.004 & 16.114 0.003 & $g$ & LT & 57029.146 & 16.844 0.125 & $u$ & LT\\ 
57129.961 & 16.106 0.003 & $g$ & LT & 57037.239 & 16.976 0.033 & $u$ & LT\\ 
57132.982 & 16.128 0.003 & $g$ & LT & 57038.213 & 17.054 0.030 & $u$ & LT\\ 
57135.962 & 16.125 0.003 & $g$ & LT & 57039.158 & 17.050 0.033 & $u$ & LT\\ 
57138.944 & 16.111 0.004 & $g$ & LT & 57041.139 & 16.970 0.044 & $u$ & LT\\ 
57142.113 & 16.152 0.008 & $g$ & LT & 57042.144 & 17.070 0.037 & $u$ & LT\\ 
57147.941 & 16.117 0.005 & $g$ & LT & 57044.154 & 17.036 0.030 & $u$ & LT\\ 
57150.951 & 16.127 0.003 & $g$ & LT & 57046.084 & 17.024 0.049 & $u$ & LT\\ 
57153.912 & 16.127 0.004 & $g$ & LT & 57048.080 & 17.034 0.045 & $u$ & LT\\ 
57158.925 & 16.105 0.004 & $g$ & LT & 57051.084 & 16.995 0.117 & $u$ & LT\\ 
57161.893 & 16.110 0.003 & $g$ & LT & 57053.057 & 17.106 0.077 & $u$ & LT\\ 
56991.426 & 15.960 0.045 & $B$ & {\swift} & 57057.083 & 17.069 0.120 & $u$ & LT\\ 
56993.882 & 16.020 0.045 & $B$ & {\swift} & 57059.044 & 16.981 0.223 & $u$ & LT\\ 
56995.275 & 15.930 0.045 & $B$ & {\swift} & 57065.161 & 17.090 0.034 & $u$ & LT\\ 
56998.228 & 15.940 0.054 & $B$ & {\swift} & 57068.132 & 17.126 0.014 & $u$ & LT\\ 
57001.602 & 16.010 0.045 & $B$ & {\swift} & 57068.144 & 17.130 0.030 & $u$ & LT\\ 
57004.132 & 15.970 0.045 & $B$ & {\swift} & 57070.128 & 17.126 0.015 & $u$ & LT\\ 
57007.260 & 16.050 0.045 & $B$ & {\swift} & 57077.224 & 17.290 0.033 & $u$ & LT\\ 
57010.798 & 16.040 0.045 & $B$ & {\swift} & 57079.073 & 17.258 0.017 & $u$ & LT\\ 
57013.061 & 16.060 0.045 & $B$ & {\swift} & 57085.034 & 17.332 0.044 & $u$ & LT\\ 
57016.056 & 16.170 0.045 & $B$ & {\swift} & 57093.078 & 17.370 0.019 & $u$ & LT\\ 
57019.510 & 16.160 0.054 & $B$ & {\swift} & 57095.073 & 17.494 0.014 & $u$ & LT\\ 
57022.712 & 16.090 0.045 & $B$ & {\swift} & 57097.996 & 17.518 0.015 & $u$ & LT\\ 
57029.315 & 16.180 0.063 & $B$ & {\swift} & 57109.929 & 17.478 0.039 & $u$ & LT\\ 
57033.110 & 16.160 0.063 & $B$ & {\swift} & 57112.019 & 17.391 0.045 & $u$ & LT\\ 
57036.045 & 16.310 0.063 & $B$ & {\swift} & 57117.979 & 17.458 0.044 & $u$ & LT\\ 
57039.032 & 16.290 0.054 & $B$ & {\swift} & 57121.023 & 17.526 0.022 & $u$ & LT\\ 
57042.228 & 16.180 0.063 & $B$ & {\swift} & 57122.961 & 17.564 0.019 & $u$ & LT\\ 
57045.555 & 16.220 0.054 & $B$ & {\swift} & 57124.977 & 17.510 0.017 & $u$ & LT\\ 
57048.752 & 16.300 0.054 & $B$ & {\swift} & 57127.006 & 17.538 0.016 & $u$ & LT\\ 
57051.465 & 16.360 0.063 & $B$ & {\swift} & 57129.963 & 17.540 0.017 & $u$ & LT\\ 
57054.004 & 16.260 0.063 & $B$ & {\swift} & 57132.984 & 17.543 0.017 & $u$ & LT\\ 
57057.524 & 16.220 0.045 & $B$ & {\swift} & 57135.965 & 17.524 0.017 & $u$ & LT\\ 
57059.796 & 16.310 0.063 & $B$ & {\swift} & 57138.946 & 17.507 0.022 & $u$ & LT\\ 
57065.779 & 16.270 0.063 & $B$ & {\swift} & 57142.115 & 17.484 0.040 & $u$ & LT\\ 
57068.775 & 16.270 0.045 & $B$ & {\swift} & 57144.907 & 17.565 0.055 & $u$ & LT\\ 
57071.704 & 16.340 0.045 & $B$ & {\swift} & 57150.953 & 17.543 0.017 & $u$ & LT\\ 
57074.844 & 16.270 0.054 & $B$ & {\swift} & 57158.927 & 17.562 0.021 & $u$ & LT\\ 
57077.572 & 16.330 0.054 & $B$ & {\swift} & 57161.895 & 17.572 0.018 & $u$ & LT\\ 
57080.889 & 16.340 0.082 & $B$ & {\swift} & 56991.425 & 15.150 0.045 & $U$ & {\swift}\\ 
57086.882 & 16.340 0.054 & $B$ & {\swift} & 56993.881 & 15.260 0.045 & $U$ & {\swift}\\ 
57089.344 & 16.340 0.045 & $B$ & {\swift} & 56995.274 & 15.220 0.045 & $U$ & {\swift}\\ 
57132.459 & 16.390 0.063 & $B$ & {\swift} & 56998.227 & 15.260 0.045 & $U$ & {\swift}\\ 
57007.258 & 16.586 0.059 & $u$ & LT & 57001.601 & 15.280 0.045 & $U$ & {\swift}\\ 
57008.281 & 16.537 0.069 & $u$ & LT & 57004.131 & 15.400 0.045 & $U$ & {\swift}\\ 
57009.276 & 16.585 0.062 & $u$ & LT & 57007.259 & 15.470 0.045 & $U$ & {\swift}\\ 
57011.228 & 16.690 0.059 & $u$ & LT & 57010.797 & 15.490 0.045 & $U$ & {\swift}\\ 
57012.212 & 16.724 0.062 & $u$ & LT & 57013.060 & 15.420 0.045 & $U$ & {\swift}\\ 
57014.176 & 16.761 0.100 & $u$ & LT & 57016.055 & 15.530 0.045 & $U$ & {\swift}\\ 
57016.214 & 16.749 0.060 & $u$ & LT & 57019.510 & 15.600 0.054 & $U$ & {\swift}\\ 
57017.206 & 16.749 0.061 & $u$ & LT & 57022.711 & 15.530 0.045 & $U$ & {\swift}\\ 
57018.164 & 16.784 0.077 & $u$ & LT & 57029.315 & 15.750 0.063 & $U$ & {\swift}\\ 
\hline
\end{tabular}
\end{minipage}
\end{table*}

\begin{table*}
\begin{minipage}{\textwidth}
\centering
\renewcommand{\arraystretch}{1.2}
\begin{tabular}{cccc|cccc}
\hline
MJD & Magnitude &  Filter & Telescope & MJD & Magnitude &  Filter & Telescope\\
\hline
57033.109 & 15.800 0.063 & $U$ & {\swift} & 57059.794 & 15.880 0.067 & $W1$ & {\swift}\\ 
57036.044 & 15.780 0.063 & $U$ & {\swift} & 57065.776 & 15.920 0.058 & $W1$ & {\swift}\\ 
57039.031 & 15.860 0.063 & $U$ & {\swift} & 57068.771 & 16.080 0.058 & $W1$ & {\swift}\\ 
57042.227 & 15.800 0.073 & $U$ & {\swift} & 57071.700 & 15.930 0.050 & $W1$ & {\swift}\\ 
57045.555 & 15.880 0.063 & $U$ & {\swift} & 57074.842 & 16.060 0.067 & $W1$ & {\swift}\\ 
57048.751 & 15.860 0.063 & $U$ & {\swift} & 57077.569 & 16.160 0.067 & $W1$ & {\swift}\\ 
57051.464 & 15.870 0.063 & $U$ & {\swift} & 57080.888 & 16.110 0.085 & $W1$ & {\swift}\\ 
57054.003 & 15.950 0.073 & $U$ & {\swift} & 57086.878 & 16.250 0.058 & $W1$ & {\swift}\\ 
57057.523 & 15.940 0.054 & $U$ & {\swift} & 57089.340 & 16.340 0.058 & $W1$ & {\swift}\\ 
57059.795 & 15.880 0.073 & $U$ & {\swift} & 57099.391 & 16.450 0.067 & $W1$ & {\swift}\\ 
57065.778 & 15.960 0.073 & $U$ & {\swift} & 57102.390 & 16.430 0.095 & $W1$ & {\swift}\\ 
57068.774 & 15.990 0.054 & $U$ & {\swift} & 57105.254 & 16.700 0.143 & $W1$ & {\swift}\\ 
57071.703 & 15.950 0.054 & $U$ & {\swift} & 57108.770 & 16.520 0.085 & $W1$ & {\swift}\\ 
57074.844 & 15.990 0.063 & $U$ & {\swift} & 57111.899 & 16.500 0.058 & $W1$ & {\swift}\\ 
57077.571 & 16.030 0.063 & $U$ & {\swift} & 57114.092 & 16.440 0.085 & $W1$ & {\swift}\\ 
57080.889 & 16.050 0.102 & $U$ & {\swift} & 57117.688 & 16.470 0.058 & $W1$ & {\swift}\\ 
57086.881 & 16.170 0.063 & $U$ & {\swift} & 57120.282 & 16.620 0.067 & $W1$ & {\swift}\\ 
57089.343 & 16.230 0.063 & $U$ & {\swift} & 57123.544 & 16.640 0.067 & $W1$ & {\swift}\\ 
57099.394 & 16.190 0.054 & $U$ & {\swift} & 57126.136 & 16.740 0.085 & $W1$ & {\swift}\\ 
57102.391 & 16.170 0.082 & $U$ & {\swift} & 57129.201 & 16.880 0.085 & $W1$ & {\swift}\\ 
57108.771 & 16.270 0.073 & $U$ & {\swift} & 57132.457 & 16.650 0.085 & $W1$ & {\swift}\\ 
57111.902 & 16.280 0.054 & $U$ & {\swift} & 57136.118 & 16.680 0.095 & $W1$ & {\swift}\\ 
57114.093 & 16.330 0.073 & $U$ & {\swift} & 57139.314 & 16.740 0.067 & $W1$ & {\swift}\\ 
57117.692 & 16.350 0.054 & $U$ & {\swift} & 57147.562 & 16.750 0.067 & $W1$ & {\swift}\\ 
57120.285 & 16.340 0.054 & $U$ & {\swift} & 57150.224 & 16.770 0.058 & $W1$ & {\swift}\\ 
57123.547 & 16.380 0.054 & $U$ & {\swift} & 57153.418 & 16.730 0.067 & $W1$ & {\swift}\\ 
57126.138 & 16.430 0.073 & $U$ & {\swift} & 57156.350 & 16.740 0.114 & $W1$ & {\swift}\\ 
57129.202 & 16.400 0.073 & $U$ & {\swift} & 56991.432 & 14.460 0.042 & $M2$ & {\swift}\\ 
57132.458 & 16.530 0.092 & $U$ & {\swift} & 56993.889 & 14.910 0.042 & $M2$ & {\swift}\\ 
57136.119 & 16.470 0.082 & $U$ & {\swift} & 56995.283 & 14.540 0.042 & $M2$ & {\swift}\\ 
57139.317 & 16.450 0.054 & $U$ & {\swift} & 56998.275 & 14.680 0.036 & $M2$ & {\swift}\\ 
57147.565 & 16.480 0.054 & $U$ & {\swift} & 57001.609 & 14.680 0.042 & $M2$ & {\swift}\\ 
57150.228 & 16.340 0.045 & $U$ & {\swift} & 57004.264 & 14.730 0.050 & $M2$ & {\swift}\\ 
57153.420 & 16.450 0.063 & $U$ & {\swift} & 57007.268 & 14.900 0.042 & $M2$ & {\swift}\\ 
57156.351 & 16.540 0.102 & $U$ & {\swift} & 57010.804 & 14.940 0.042 & $M2$ & {\swift}\\ 
56991.423 & 14.730 0.042 & $W1$ & {\swift} & 57013.068 & 14.950 0.042 & $M2$ & {\swift}\\ 
56993.878 & 14.980 0.042 & $W1$ & {\swift} & 57016.063 & 15.140 0.042 & $M2$ & {\swift}\\ 
56995.271 & 14.820 0.042 & $W1$ & {\swift} & 57019.516 & 15.170 0.042 & $M2$ & {\swift}\\ 
56998.225 & 14.880 0.050 & $W1$ & {\swift} & 57022.719 & 15.180 0.042 & $M2$ & {\swift}\\ 
57001.598 & 14.890 0.042 & $W1$ & {\swift} & 57029.319 & 15.580 0.050 & $M2$ & {\swift}\\ 
57004.128 & 14.940 0.042 & $W1$ & {\swift} & 57033.114 & 15.370 0.050 & $M2$ & {\swift}\\ 
57007.256 & 15.150 0.042 & $W1$ & {\swift} & 57036.109 & 15.490 0.050 & $M2$ & {\swift}\\ 
57010.794 & 15.240 0.042 & $W1$ & {\swift} & 57039.037 & 15.500 0.050 & $M2$ & {\swift}\\ 
57013.057 & 15.210 0.042 & $W1$ & {\swift} & 57042.231 & 15.620 0.058 & $M2$ & {\swift}\\ 
57016.052 & 15.340 0.050 & $W1$ & {\swift} & 57045.561 & 15.620 0.050 & $M2$ & {\swift}\\ 
57019.507 & 15.400 0.050 & $W1$ & {\swift} & 57051.469 & 15.660 0.050 & $M2$ & {\swift}\\ 
57022.708 & 15.390 0.050 & $W1$ & {\swift} & 57054.007 & 15.670 0.050 & $M2$ & {\swift}\\ 
57029.313 & 15.620 0.058 & $W1$ & {\swift} & 57057.531 & 15.740 0.050 & $M2$ & {\swift}\\ 
57033.038 & 15.690 0.058 & $W1$ & {\swift} & 57059.799 & 15.780 0.058 & $M2$ & {\swift}\\ 
57036.042 & 15.640 0.050 & $W1$ & {\swift} & 57065.853 & 15.700 0.076 & $M2$ & {\swift}\\ 
57039.029 & 15.730 0.050 & $W1$ & {\swift} & 57068.782 & 16.070 0.050 & $M2$ & {\swift}\\ 
57042.226 & 15.750 0.067 & $W1$ & {\swift} & 57071.711 & 15.870 0.050 & $M2$ & {\swift}\\ 
57045.552 & 15.820 0.058 & $W1$ & {\swift} & 57074.848 & 15.950 0.058 & $M2$ & {\swift}\\ 
57051.463 & 15.780 0.058 & $W1$ & {\swift} & 57077.577 & 16.020 0.050 & $M2$ & {\swift}\\ 
57054.002 & 15.780 0.067 & $W1$ & {\swift} & 57080.891 & 16.120 0.076 & $M2$ & {\swift}\\ 
57057.520 & 15.770 0.050 & $W1$ & {\swift} & 57086.888 & 16.130 0.050 & $M2$ & {\swift}\\ 
\hline
\end{tabular}
\end{minipage}
\end{table*}

\begin{table*}
\begin{minipage}{\textwidth}
\centering
\renewcommand{\arraystretch}{1.2}
\begin{tabular}{cccc|cccc}
\hline
MJD & Magnitude &  Filter & Telescope & MJD & Magnitude &  Filter & Telescope\\
\hline
57089.351 & 16.300 0.050 & $M2$ & {\swift} & 57036.046 & 15.560 0.058 & $W2$ & {\swift}\\ 
57099.388 & 16.310 0.067 & $M2$ & {\swift} & 57039.033 & 15.350 0.042 & $W2$ & {\swift}\\ 
57102.389 & 16.600 0.104 & $M2$ & {\swift} & 57042.228 & 15.400 0.050 & $W2$ & {\swift}\\ 
57105.250 & 16.450 0.058 & $M2$ & {\swift} & 57045.556 & 15.510 0.050 & $W2$ & {\swift}\\ 
57108.768 & 16.490 0.085 & $M2$ & {\swift} & 57051.466 & 15.440 0.050 & $W2$ & {\swift}\\ 
57111.895 & 16.580 0.067 & $M2$ & {\swift} & 57054.004 & 15.550 0.050 & $W2$ & {\swift}\\ 
57114.090 & 16.570 0.095 & $M2$ & {\swift} & 57057.525 & 15.570 0.042 & $W2$ & {\swift}\\ 
57117.683 & 16.560 0.058 & $M2$ & {\swift} & 57059.796 & 15.580 0.050 & $W2$ & {\swift}\\ 
57120.278 & 16.700 0.067 & $M2$ & {\swift} & 57065.779 & 15.550 0.085 & $W2$ & {\swift}\\ 
57123.540 & 16.670 0.067 & $M2$ & {\swift} & 57068.776 & 16.050 0.050 & $W2$ & {\swift}\\ 
57126.134 & 16.770 0.085 & $M2$ & {\swift} & 57071.705 & 15.740 0.042 & $W2$ & {\swift}\\ 
57129.198 & 16.810 0.085 & $M2$ & {\swift} & 57074.845 & 15.770 0.050 & $W2$ & {\swift}\\ 
57132.462 & 16.740 0.076 & $M2$ & {\swift} & 57077.573 & 15.800 0.050 & $W2$ & {\swift}\\ 
57136.116 & 16.790 0.104 & $M2$ & {\swift} & 57080.890 & 15.940 0.067 & $W2$ & {\swift}\\ 
57139.309 & 16.670 0.067 & $M2$ & {\swift} & 57086.883 & 16.030 0.050 & $W2$ & {\swift}\\ 
57147.558 & 16.790 0.067 & $M2$ & {\swift} & 57089.345 & 16.120 0.050 & $W2$ & {\swift}\\ 
57150.218 & 16.850 0.067 & $M2$ & {\swift} & 57099.395 & 16.250 0.050 & $W2$ & {\swift}\\ 
57153.414 & 16.750 0.076 & $M2$ & {\swift} & 57102.392 & 16.360 0.067 & $W2$ & {\swift}\\ 
57156.349 & 16.860 0.124 & $M2$ & {\swift} & 57108.772 & 16.310 0.067 & $W2$ & {\swift}\\ 
56991.427 & 14.260 0.042 & $W2$ & {\swift} & 57111.904 & 16.450 0.050 & $W2$ & {\swift}\\ 
56993.883 & 14.700 0.042 & $W2$ & {\swift} & 57114.094 & 16.350 0.067 & $W2$ & {\swift}\\ 
56995.276 & 14.310 0.042 & $W2$ & {\swift} & 57117.694 & 16.480 0.058 & $W2$ & {\swift}\\ 
56998.229 & 14.360 0.050 & $W2$ & {\swift} & 57120.286 & 16.690 0.058 & $W2$ & {\swift}\\ 
57001.603 & 14.460 0.042 & $W2$ & {\swift} & 57123.549 & 16.630 0.058 & $W2$ & {\swift}\\ 
57004.133 & 14.480 0.042 & $W2$ & {\swift} & 57126.139 & 16.760 0.067 & $W2$ & {\swift}\\ 
57007.261 & 14.690 0.042 & $W2$ & {\swift} & 57129.203 & 16.830 0.076 & $W2$ & {\swift}\\ 
57010.799 & 14.770 0.042 & $W2$ & {\swift} & 57132.459 & 16.660 0.067 & $W2$ & {\swift}\\ 
57013.062 & 14.730 0.042 & $W2$ & {\swift} & 57136.120 & 16.660 0.076 & $W2$ & {\swift}\\ 
57016.057 & 14.920 0.042 & $W2$ & {\swift} & 57139.318 & 16.590 0.050 & $W2$ & {\swift}\\ 
57019.511 & 15.010 0.042 & $W2$ & {\swift} & 57147.567 & 16.670 0.050 & $W2$ & {\swift}\\ 
57022.713 & 14.990 0.042 & $W2$ & {\swift} & 57150.230 & 16.720 0.050 & $W2$ & {\swift}\\ 
57029.316 & 15.380 0.050 & $W2$ & {\swift} & 57153.422 & 16.790 0.058 & $W2$ & {\swift}\\ 
57033.111 & 15.170 0.050 & $W2$ & {\swift} & 57156.352 & 16.680 0.085 & $W2$ & {\swift}\\
\hline
\end{tabular}

\medskip
\raggedright
\noindent Magnitudes and uncertainties are presented in the natural system for each filter: $ugriz$ magnitudes are presented in the AB system, {\swift} filter magnitudes are presented in the Vega system. Uncertainties are given next to the magnitude measurements. Data are not corrected for Galactic extinction.
\end{minipage}
\end{table*}


\begin{table*}
\begin{minipage}{\textwidth}
\centering
\caption{Swift XRT photometry of {\name}.}
\renewcommand{\arraystretch}{1.2}
\begin{tabular}{ccccc|ccccc}
\hline
MJD & Flux  & Uncertainty & \edit{Count Rate} & \edit{Uncertainty} & MJD & Flux  & Uncertainty & \edit{Count Rate} & \edit{Uncertainty} \\
\hline
56991.5 & 1.81 & 0.10 & \edit{0.35} & \edit{0.01} & 57068.5 & 1.60 & 0.12 & \edit{0.24} & \edit{0.01}\\ 
56993.5 & 1.89 & 0.10 & \edit{0.33} & \edit{0.01} & 57071.5 & 1.64 & 0.10 & \edit{0.25} & \edit{0.01}\\ 
56995.5 & 2.09 & 0.09 & \edit{0.33} & \edit{0.01} & 57074.5 & 1.68 & 0.10 & \edit{0.24} & \edit{0.01}\\ 
56998.5 & 1.92 & 0.10 & \edit{0.41} & \edit{0.02} & 57077.5 & 1.56 & 0.15 & \edit{0.19} & \edit{0.01}\\ 
57001.5 & 2.01 & 0.09 & \edit{0.39} & \edit{0.02} & 57080.5 & 1.85 & 0.10 & \edit{0.24} & \edit{0.01}\\ 
57004.5 & 2.17 & 0.18 & \edit{0.41} & \edit{0.02} & 57086.9 & 1.81 & 0.10 & \edit{0.27} & \edit{0.01}\\ 
57007.5 & 2.60 & 0.09 & \edit{0.45} & \edit{0.02} & 57089.3 & 1.58 & 0.08 & \edit{0.30} & \edit{0.01}\\ 
57010.5 & 2.60 & 0.11 & \edit{0.48} & \edit{0.02} & 57099.4 & 2.11 & 0.17 & \edit{0.25} & \edit{0.01}\\ 
57013.5 & 2.27 & 0.09 & \edit{0.43} & \edit{0.02} & 57102.6 & 1.71 & 0.10 & \edit{0.23} & \edit{0.01}\\ 
57016.5 & 1.98 & 0.09 & \edit{0.45} & \edit{0.02} & 57109.2 & 1.77 & 0.12 & \edit{0.23} & \edit{0.01}\\ 
57019.5 & 2.17 & 0.07 & \edit{0.40} & \edit{0.01} & 57111.9 & 2.34 & 0.11 & \edit{0.23} & \edit{0.01}\\ 
57022.5 & 2.36 & 0.11 & \edit{0.33} & \edit{0.01} & 57114.1 & 1.48 & 0.10 & \edit{0.20} & \edit{0.01}\\ 
57029.5 & 2.22 & 0.07 & \edit{0.40} & \edit{0.01} & 57117.7 & 2.03 & 0.13 & \edit{0.20} & \edit{0.01}\\ 
57033.5 & 2.29 & 0.11 & \edit{0.36} & \edit{0.02} & 57120.3 & 1.51 & 0.08 & \edit{0.19} & \edit{0.01}\\ 
57037.5 & 2.50 & 0.11 & \edit{0.39} & \edit{0.02} & 57123.6 & 1.70 & 0.13 & \edit{0.19} & \edit{0.01}\\ 
57040.5 & 2.01 & 0.12 & \edit{0.35} & \edit{0.01} & 57126.2 & 1.53 & 0.10 & \edit{0.17} & \edit{0.01}\\ 
57043.5 & 1.89 & 0.09 & \edit{0.36} & \edit{0.02} & 57129.4 & 1.62 & 0.13 & \edit{0.18} & \edit{0.01}\\ 
57046.5 & 1.99 & 0.11 & \edit{0.36} & \edit{0.02} & 57132.6 & 1.08 & 0.09 & \edit{0.16} & \edit{0.01}\\ 
57048.5 & 1.91 & 0.09 & \edit{0.33} & \edit{0.01} & 57136.6 & 1.05 & 0.07 & \edit{0.15} & \edit{0.01}\\ 
57051.5 & 2.00 & 0.09 & \edit{0.30} & \edit{0.02} & 57139.3 & 1.38 & 0.13 & \edit{0.17} & \edit{0.01}\\ 
57053.5 & 1.77 & 0.09 & \edit{0.31} & \edit{0.01} & 57147.6 & 1.05 & 0.08 & \edit{0.14} & \edit{0.01}\\ 
57057.5 & 1.80 & 0.09 & \edit{0.28} & \edit{0.01} & 57150.3 & 1.72 & 0.14 & \edit{0.15} & \edit{0.01}\\ 
57059.5 & 1.65 & 0.08 & \edit{0.29} & \edit{0.01} & 57153.5 & 1.39 & 0.10 & \edit{0.15} & \edit{0.01}\\ 
57065.5 & 1.66 & 0.10 & \edit{0.20} & \edit{0.01} & 57156.7 & 1.18 & 0.11 & \edit{0.15} & \edit{0.01}\\ 
\hline
\end{tabular}

\medskip
\raggedright
\noindent \edit{All X-ray fluxes and flux uncertainties are given in units of $10^{-11}$ ergs~s$^{-1}$~cm$^{-2}$ while count rates and count rate uncertainties are given in counts s$^{-1}$. The Swift XRT energy range is $0.3-10$ keV.} Data are not corrected for Galactic extinction.
\label{tab:xray}
\end{minipage}
\end{table*}


\begin{table*}
\begin{minipage}{\textwidth}
\centering
\caption{Spectroscopic Observations of {\name}.}
\renewcommand{\arraystretch}{1.2}
\begin{tabular}{lclr}
\hline
\multicolumn{1}{c}{UT Date} & MJD & \multicolumn{1}{c}{Telescope/Instrument} & \multicolumn{1}{c}{Exposure (s)} \\
\hline
2014 November 30.60 & 56991.60 & UH-2.2m/SNIFS & $2\times400$ \\
2014 December 02.51 & 56993.51 & APO-3.5m/DIS & $2\times1000$ \\
2014 December 09.51 & 57000.51 & MDM-2.4m/Modspec & $4\times600$ \\
2014 December 10.56 & 57001.56 & MDM-2.4m/Modspec & $3\times240$ \\
2014 December 12.48 & 57003.48 & APO-3.5m/DIS & $2\times1200$ \\
2014 December 12.54 & 57003.54 & MDM-2.4m/Modspec & $1\times600$ \\
2014 December 14.49 & 57005.49 & FLWO-1.5m/FAST & $3\times900$ \\
2014 December 15.51 & 57006.51 & FLWO-1.5m/FAST & $3\times1200$ \\
2014 December 19.49 & 57010.49 & FLWO-1.5m/FAST & $1\times1800$ \\
2014 December 23.51 & 57014.51 & MDM-2.4m/OSMOS & $3\times1200$ \\
2015 January 03.35 &  57025.35 & Magellan-6.5m/IMACS & $2\times500$ \\
2015 January 20.53 & 57042.53 & LBT-8.4m/MODS & $2\times900$ \\ 
2015 February 11.40 & 57064.40 & MDM-2.4m/OSMOS & $3\times1200$ \\
2015 February 14.50 & 57067.50 & FLWO-1.5m/FAST & $2\times1200$ \\
2015 February 15.38 & 57069.38 & LBT-8.4m/MODS & $3\times1200$ \\
2015 March 11.35 & 57092.35 & MDM-2.4m/OSMOS & $3\times1200$ \\
2015 April 17.19 & 57129.19 & LBT-8.4m/MODS & $3\times900$ \\
\hline
\end{tabular}
\label{tab:spectra}
\end{minipage}
\end{table*}


\begin{table*}
\begin{minipage}{\textwidth}
\centering
\caption{Line luminosities for {\name}.}
\renewcommand{\arraystretch}{1.2}
\begin{tabular}{cccc}
\hline
MJD & H$\alpha$ (6563~\AA) & H$\beta$ (4861~\AA) & \ion{He}{2} (4686~\AA) \\
\hline
56993.51 & $1.01\times10^{41}$ & $3.85\times10^{40}$ & $5.07\times10^{40}$ \\
57001.56 & $1.01\times10^{41}$ & $2.98\times10^{40}$ & $5.53\times10^{40}$ \\
57003.48 & $1.12\times10^{41}$ & $3.65\times10^{40}$ & $7.09\times10^{40}$ \\
57003.54 & $8.27\times10^{40}$ & $2.80\times10^{40}$ & $5.45\times10^{40}$ \\
57005.49 & $9.72\times10^{40}$ & $3.65\times10^{40}$ & $6.06\times10^{40}$ \\
57006.51 & $7.69\times10^{40}$ & $5.63\times10^{40}$ & $8.67\times10^{40}$ \\
57010.49 & $9.86\times10^{40}$ & $2.82\times10^{40}$ & $6.52\times10^{40}$ \\
57014.51 & $6.72\times10^{40}$ & $2.64\times10^{40}$ & $3.12\times10^{40}$ \\
57025.35 & $6.58\times10^{40}$ & $1.99\times10^{40}$ & $3.77\times10^{40}$ \\
57042.53 & $4.90\times10^{40}$ & $1.91\times10^{40}$ & $5.07\times10^{40}$ \\
57064.40 & $2.10\times10^{40}$ & $1.29\times10^{40}$ & $3.79\times10^{40}$ \\
57067.49 & $2.55\times10^{40}$ & $1.42\times10^{40}$ & $2.38\times10^{40}$ \\
57069.38 & $4.07\times10^{40}$ & $1.16\times10^{40}$ & $2.91\times10^{40}$ \\
57092.35 & $1.41\times10^{40}$ & --- & $2.05\times10^{40}$ \\
57129.19 & $1.59\times10^{40}$ & $6.96\times10^{39}$ & $1.70\times10^{40}$ \\
\hline
\end{tabular}

\medskip
\raggedright
\noindent Column headers indicate the lines used to measure the luminosities. All luminosities are given in units of ergs s$^{-1}$. No value is given for epochs where a measurement was not possible.
\label{tab:line_lum}
\end{minipage}
\end{table*}

\end{document}